\documentclass[english,elsarticle]{elsarticle}
\usepackage[T1]{fontenc}
\usepackage[utf8]{inputenc}
\usepackage{geometry}
\usepackage{color}
\usepackage{url}
\usepackage{amstext}
\usepackage{graphicx}
\usepackage{amssymb}

\makeatletter
\journal{Comptes Rendus de l'Académie des Sciences: C. R. Physique 13 (2012) 89–100}

\makeatother

\usepackage{babel}

\begin{document}

\title{Superconducting Quantum Point Contacts}

\author{L. Bretheau}

\author{Ç. Girit}

\author{L. Tosi}

\author{M. Goffman}

\author{P. Joyez}

\author{H. Pothier}

\author{D. Esteve}

\author{C. Urbina\corref{cor1}}

\ead{cristian.urbina@cea.fr}

\cortext[cor1]{Corresponding author}

\address{Quantronics Group, Service de Physique de l'État Condensé (CNRS,
URA\ 2464), IRAMIS, CEA-Saclay, 91191 Gif-sur-Yvette, France}
\begin{abstract}
We review our experiments on the electronic transport properties of
atomic contacts between metallic electrodes, in particular superconducting
ones. Despite ignorance of the exact atomic configuration, these ultimate
quantum point contacts can be manipulated and well characterized in-situ.
They allow performing fundamental tests of the scattering theory of
quantum transport. In particular, we discuss the case of the Josephson
effect.
\end{abstract}
\begin{keyword}
atomic contacts \sep superconductivity \sep quantum transport \PACS
74.45.+c \sep 74.50.+r \sep 73.23.-b 
\end{keyword}
\maketitle

\section{Introduction}

For significant quantum effects to appear in the transport properties
of a conductor, it must be shorter than the distance over which an
electron propagates without losing its quantum coherence, which for
usual metals at low temperatures (<1K) is on the order of microns.
The exploration of this quantum transport regime started in the early
1980s with the observation of Aharonov-Bohm interferences in submicron
metallic rings \citep{webb_observation_1985}. A very powerful point
of view, put forward by R. Landauer \citep{landauer_electrical_1970,martin_wave-packet_1992},
describes a transport experiment across a coherent conductor as the
scattering of the quantum electronic waves injected by the contact
probes. In other words, the conductor is viewed as an electronic waveguide,
whose modes, called {}``conduction channels'', are each characterized
by a transmission probability. For example, the total conductance
is given by the famous Landauer formula: $G=G_{0}\sum_{1}^{N}\tau_{i}$,
where $N$ is the number of channels, the $\tau_{i}$ are the individual
transmission probabilities, and $G_{0}=2e^{2}/h\sim77\mu S$ the conductance
quantum. In fact, in systems with non-interacting quasiparticles all
transport quantities can be expressed in terms of the transmission
probability set $\left\{ \tau_{i}\right\} $, which is viewed in this
framework as the “Personal Identity Number'' (PIN code) of the conductor
\citep{nazarov_quantum_2009}. For most systems, the number of conduction
channels is very large, and the description of transport can only
be based on a distribution function for the transmission eigenvalues.
However, the number of channels can be significantly reduced if the
lateral dimensions of a conductor become comparable to the Fermi wavelength
of the electrons. This was demonstrated in 1988 using constrictions
tailored in the two-dimensional electron gas of a semiconductor heterostructure,
with a width adjustable on the scale of the Fermi wavelength ($\sim50{\normalcolor \mathrm{\, nm}}$)
by means of an electrostatic gate \citep{van_wees_quantized_1988,wharam_one-dimensional_1988}.
In these {}``quantum point contacts'', as the constriction width
is continuously decreased, a descending staircase conductance curve
is observed, beautifully illustrating the closing of the channels
one after another. Since then much activity has been devoted worldwide
to the investigation of transport through a broad range of coherent
conductors (diffusive and ballistic conductors, quantum dots, nanowires,
individual molecules like carbon nanotubes) connected to reservoirs
in the normal or in the superconducting state \citep{nazarov_quantum_2009}. 

Among the various systems investigated, atomic-size contacts \citep{agrait_quantum_2003}
between three-dimensional metallic electrodes play an important role.
These contacts have been accessible since the early years of STM \citep{eigler_atomic_1991},
and more stable ones were achieved at the beginning of the 1990s by
the group of Jan van Ruitenbeek at Leiden using the break-junction
technique \citep{moreland_electron_1985,muller_experimental_1992}.
Since their characteristic transverse dimensions are of the order
of the Fermi wavelength (typically $0.2\,\mathrm{nm}$), atomic contacts
accommodate only a small number of channels and behave as quantum
point contacts even at room temperature. An interesting difference
with respect to quantum point contacts in 2D electron systems is that
superconductivity can be built-in with the proper choice of material.
The discovery by our group in 1997 that the PIN-code could be accurately
measured for every contact \citep{scheer_conduction_1997} paved the
way to a new generation of quantum transport experiments in which
the measured transport quantities could be compared to the theoretical
predictions with no adjustable parameters. As an example, we describe
here our experiments on the supercurrent flowing through atomic-size
contacts. They probe the main concepts of the {}``mesoscopic'' theory
of the Josephson effect, \emph{i.e.} the theory in terms of the transmission
probability of the channels, that we sketch in section 3. Before discussing
previous and on-going experiments in sections 4 and 5 respectively,
we describe the microfabrication technique that has made possible
the experiments and the procedure we use to determine the PIN-code
of atomic-size contacts in section 2.

\section{Production and characterization of atomic contacts}

\subsection{Microfabricated break junctions}

\begin{figure}[tbh]
\begin{centering}
\includegraphics[height=130pt]{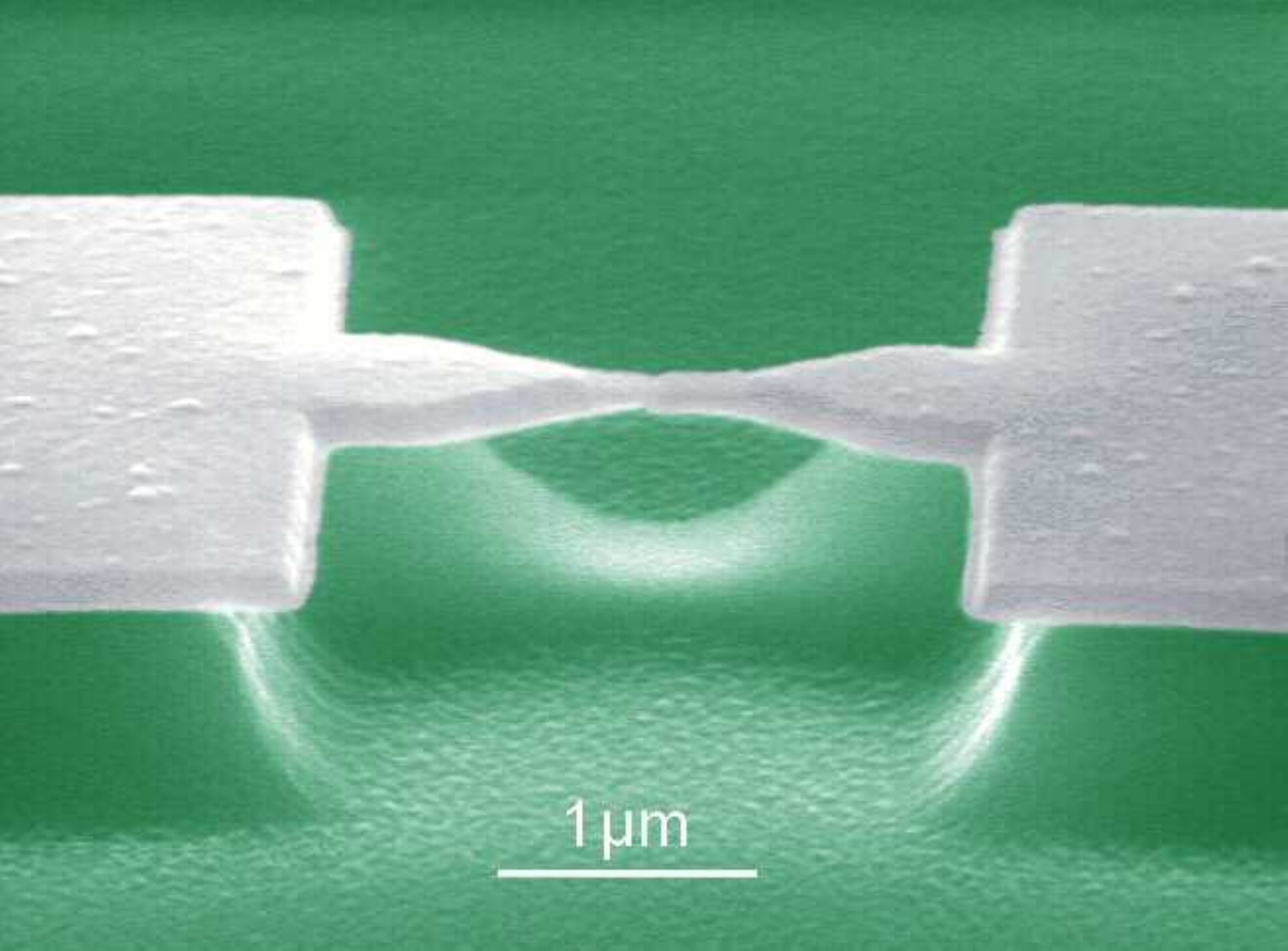}\includegraphics[height=130pt]{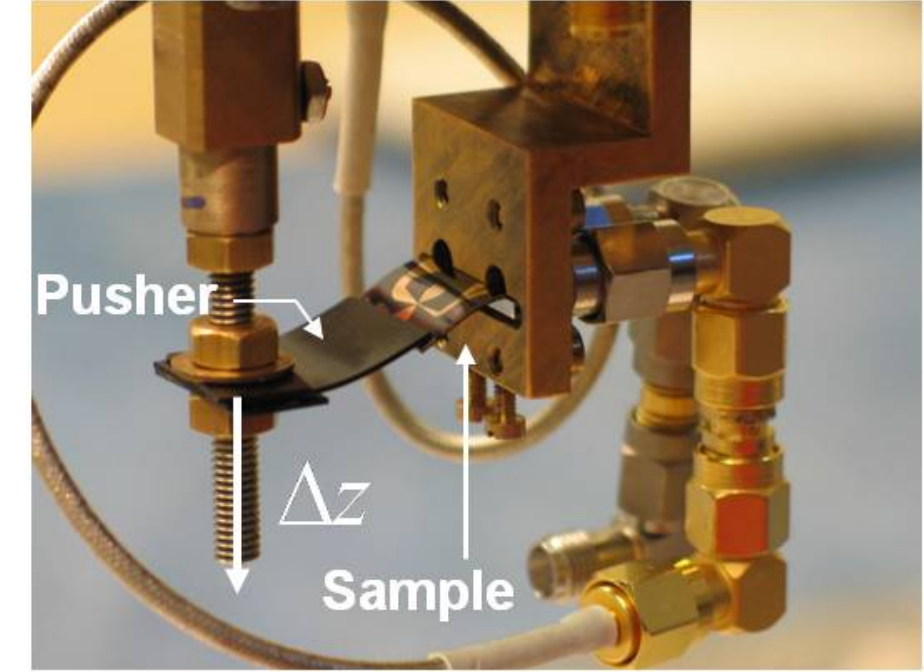}\includegraphics[height=130pt]{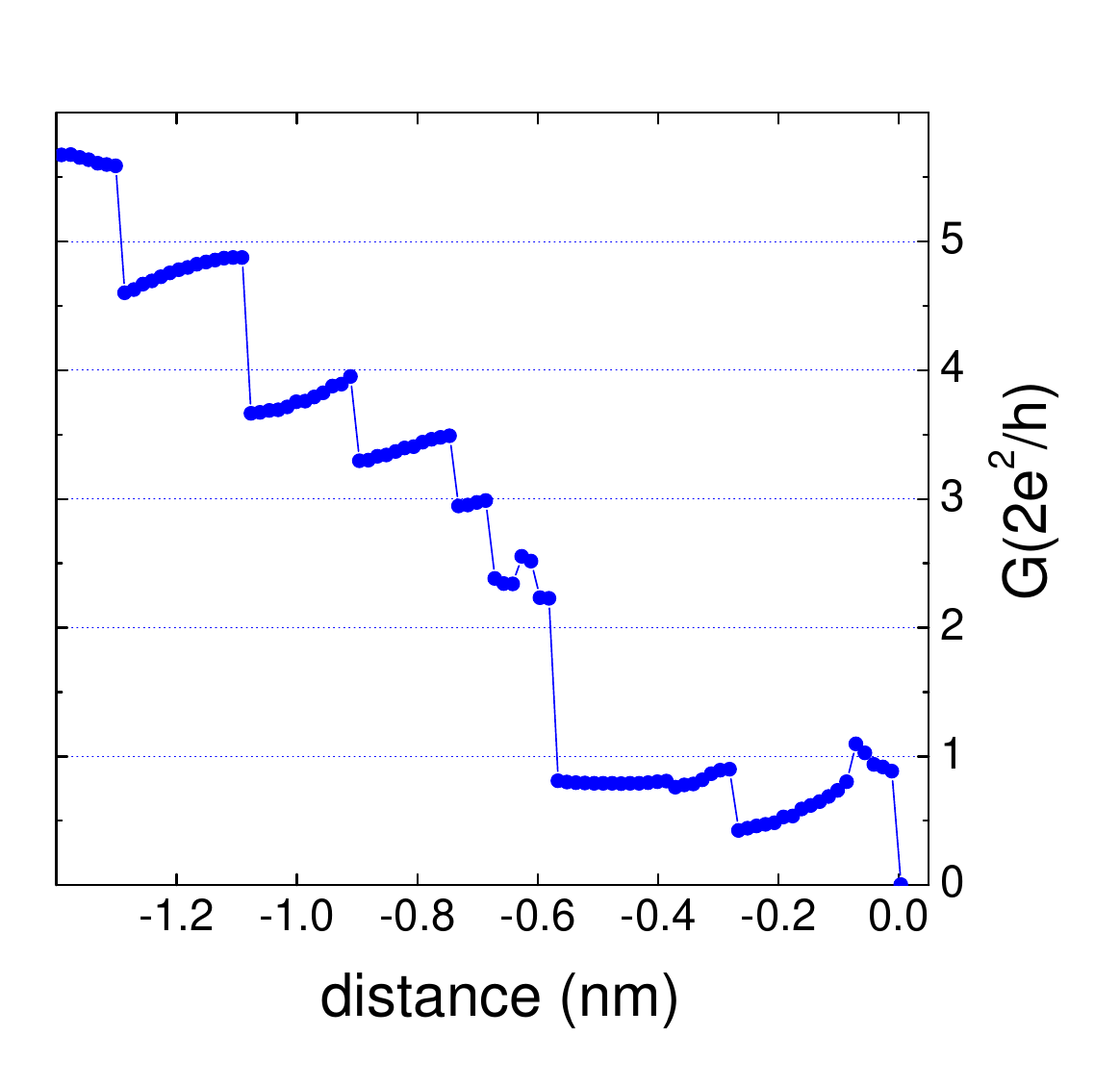}
\par\end{centering}

\caption{Principle of the microfabricated break-junction technique.\textbf{
Left:} A thin microfabricated aluminum bridge is broken by bending
the substrate on which it is fabricated. \textbf{Center:} Example
of a bending mechanism allowing microwave measurements. The ensemble
is placed in cryogenic vacuum and cooled below 100mK. A rod, driven
by a room temperature dc motor, pushes the free end of the sample
which is firmly clamped on the opposite side against two microwave
SMA launchers. Typically, a vertical displacement $\Delta z=1\,\mathrm{\mu m}$
of the pushing rod results in a $10-100\,\mathrm{pm}$ change of the
distance between the electrodes of the bridge. \textbf{Right: }Conductance
of a bridge, in units of the conductance quantum, as a function of
displacement when pulling the electrodes apart. The origin of the
displacement was set arbitrarily at the point where the bridge actually
breaks and the conductance starts decreasing exponentially signaling
the tunnel regime. Measurements were performed at 50 mK under a magnetic
field of 200 mT to suppress superconductivity in the aluminum films. }
\label{Fig:MBJ}%
\end{figure}

In order to produce atomic contacts we developed the microfabricated
break-junction (MBJ) technique \citep{van_ruitenbeek_adjustable_1996}.
Using electron beam lithography and reactive ion etching, a metallic
bridge clamped to an elastic substrate is suspended over a few microns
between two anchors (see left panel of Fig.~\ref{Fig:MBJ}). The
bridge has a constriction at the center with a diameter of approximately
100 nm. Using the mechanism shown in the center panel of Fig.~\ref{Fig:MBJ}
the substrate is first bent until the bridge breaks at the constriction.
This is performed under cryogenic vacuum so that there is no contamination
of the two resulting electrodes which are then gently brought back
into contact. Afterward, the contact can be repeatedly made and broken
at will.

The right panel of Fig.~\ref{Fig:MBJ} displays a typical conductance
evolution observed while opening an aluminum MBJ. The conductance
evolves through a series of slanted plateaus and sharp steps. The
conductance on the last plateau before full opening and the height
of the previous steps is of the order of the conductance quantum $G_{0}$.
Although similar staircase patterns are observed every time the experiment
is performed, the horizontal extension (of the order of a fraction
of nanometer) and the vertical position of the plateaus are not always
the same for subsequent compression-extension cycles. These general
features have been observed by many groups for a variety of metals
under different experimental conditions (temperature, technique used
to produce atomic size contacts, rate of compression and extension).
The typical conductance on the last plateau, the typical lengths of
the plateaus, and the behavior within the plateaus are characteristic
of each material \citep{agrait_quantum_2003}.

The succession of plateaus and conductance jumps is directly related
to the dynamics of the atomic configuration of the contact. Combined
STM-AFM experiments that measure the force between the tip and the
surface simultaneously with the conductance have directly shown that
on a plateau the atomic configuration is only elastically deformed
while a conductance jump results from an abrupt reconfiguration of
the atoms at the contact accompanied by stress relief \citep{rubio_atomic-sized_1996,rubio-bollinger_metallic_2004}.
Molecular dynamics simulations \citep{todorov_jumps_1993,brandbyge_quantized_1995,sorensen_mechanical_1998}
confirm this interpretation of the staircase pattern. These were the
first clues that the smallest contacts are indeed made of a single
atom. In the case of gold, one-atom contacts and atomic chains \citep{yanson_formation_1998}
have even been observed directly with high resolution Transmission
Electron Microscopes \citep{ohnishi_quantized_1998,rodrigues_real-time_2001,masuda__2009}.

As compared to other techniques, microfabricated break junctions present
several major advantages essential to the realization of the experiments
presented here. First, atomic-size contacts fabricated this way are
extremely stable and can be maintained in the same configuration for
weeks. Second, with a given suspended bridge many different contacts
can be created in-situ, allowing exploration of the physics of interest
for a broad range of transmission coefficients. Finally, the flexibility
offered by microfabrication techniques makes possible embedding the
contacts into on-chip circuits.

\subsection{Determination of the PIN code}

\begin{figure}[tbh]
\begin{centering}
\includegraphics[height=140pt]{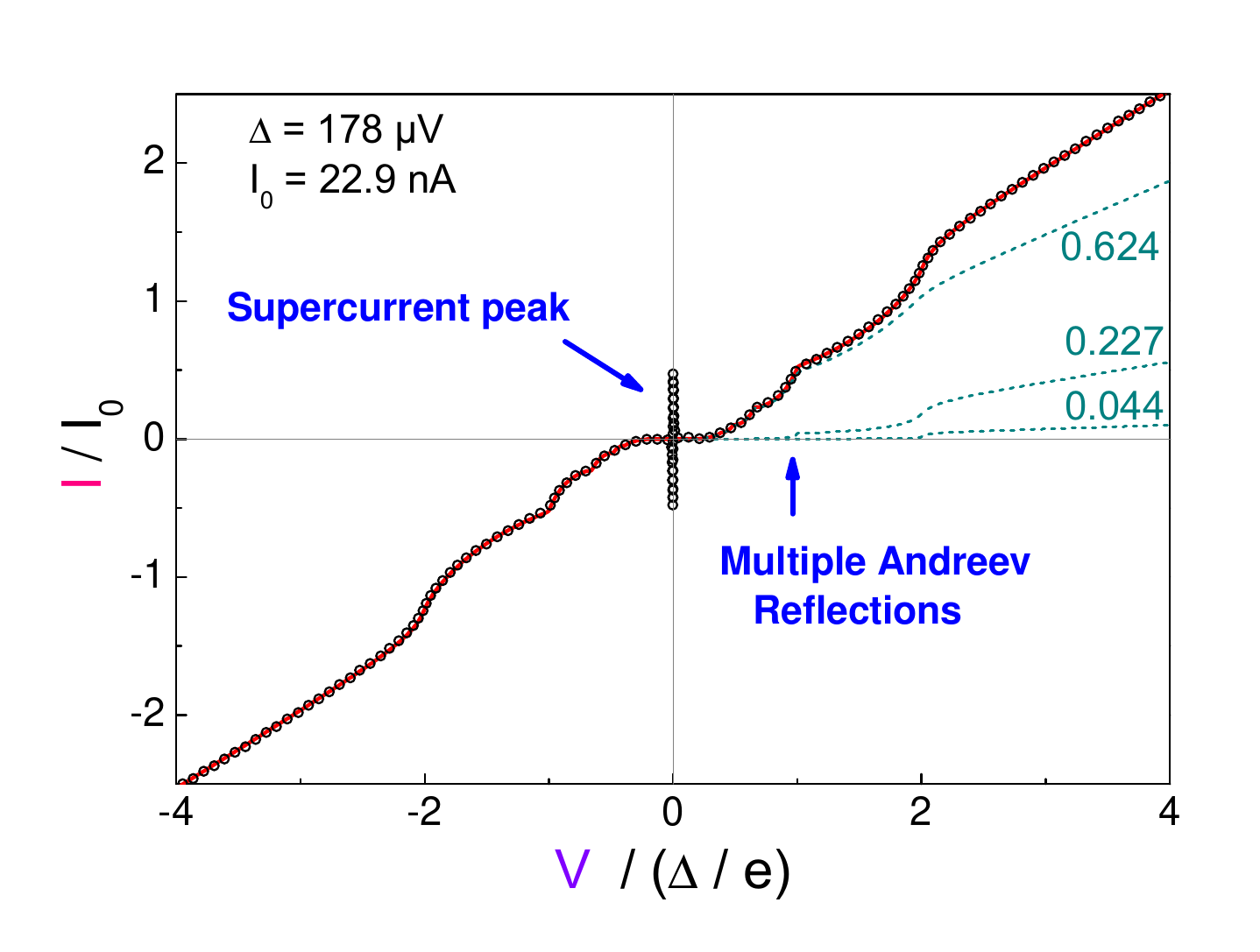}
\par\end{centering}

\caption{$I(V)$ characteristics of an atomic contact. \textbf{Open symbols}:
experimental data. \textbf{Full red line}: best fit resulting from
the sum of three single channel characteristics \textbf{(dashed lines)}
calculated using the theory of MAR and the transmission values indicated
on the right edge.\label{fig:FitMAR}}
\end{figure}
Our first major result was devising a method to determine not only
the number of channels accommodated by a contact but also its PIN
code with good accuracy \citep{scheer_conduction_1997,cron_atomic_2001}.
This was achieved by measuring the current-voltage characteristic
$I\left(V\right)$ of the contact in the superconducting state (see
Fig.~\ref{fig:FitMAR})%
\footnote{\textcolor{black}{Current-noise measurements in the normal state can
also be used to determine the PIN code, but only for contacts containing
no more than two channels}\textcolor{red}{{} }\textcolor{black}{\citep{agrait_quantum_2003}.}%
}. At exactly zero voltage a finite current,\emph{ i.e.} a supercurrent,
can flow. Above some bias current, the system switches to a finite
voltage state, \emph{i.e. }the current becomes dissipative\emph{.}
The voltage and transmission dependence of the dissipative current
through a weak link was explained for the first time in the early
1980s in terms of Multiple Andreev Reflections (MAR) \citep{klapwijk_explanation_1982,blonder_transition_1982}.
A full quantum theory of MAR for a single channel of arbitrary transmission
was achieved in the mid 90s \citep{arnold_superconducting_1987,averin_ac_1995,cuevas_hamiltonian_1996}.
Three numerically calculated current-voltage characteristics resulting
from these MAR processes are shown in Fig.~\ref{fig:FitMAR}. These
elementary curves are highly non-linear below twice the superconducting
gap $\Delta$ and present current steps at voltages $2\Delta/en$,
with $n$ an integer, which mark the onset of MAR of different orders.
The n-th order process involves the transfer of $n$ electrons, and
in a given channel its intensity varies as the n-th power of the transmission.
Consequently, the $I(V)$ characteristic depends on all powers of
every transmission coefficient and carries sufficient information
to reconstruct the PIN code. The determination of the PIN code of
any atomic-size contact is achieved by decomposing the measured total
current into a series of such elementary characteristics, each of
them corresponding to a well defined transmission probability. The
individual transmission probabilities of the channels are adjusted
so as to get the best fit of the measured current-voltage characteristic
\citep{riquelme_distribution_2005}. This automatically yields the
number of channels having a non-negligible transmission. The uncertainty
on the transmissions obtained by this fitting procedure depends on
the uncertainty in the measurement of both voltage and current, on
the transmission of the channels participating in the contact, and
on the environmental impedance. Typically, for channels with a large
transmission the relative uncertainty achieved on $\tau_{i}$ is of
the order of $0.1\%$ %
\footnote{for sufficiently low environment impedance to avoid significant Coulomb
blockade effects.%
} For channels having low transmissions $\left(\tau<.05\right)$, the
procedure fails to disentangle the contributions of the different
channels, and yields a larger error bar (see Chapter~1 of \citep{cron_atomic_2001}
for details).

Using this method we showed that for one-atom contacts the number
of conduction channels is directly related to the number of valence
orbitals of the metal \citep{scheer_signature_1998}. For aluminum
and lead, which have p-electrons at the Fermi level, three channels
were found to contribute to the conductance of the smallest contacts.
For Niobium, a transition metal with s- and d-electrons, five conduction
channels were found. In the case of gold, with one s-electron, and
for which superconductivity was induced through the proximity effect
by intimate contact with a thick aluminum layer, we found that only
one conduction channel contributes to the conductance on the lowest
conductance plateau \citep{scheer_proximity_2001}.

\section{Mesoscopic theory of the Josephson effect}

In 1962, Josephson predicted that a dc current could flow at zero
bias voltage between two superconducting electrodes coupled by a tunnel
barrier \citep{josephson_possible_1962}. This supercurrent results
from the coherent transfer of Cooper pairs and is driven by the phase
difference $\delta=\delta_{L}-\delta_{R}$ between the superconducting
order parameters of the two electrodes. The phase difference is related
to the voltage difference $V$ between the electrodes by the Josephson
relation $\dot{\delta}=2\pi V/\phi_{0}$, with $\phi_{0}=h/2e$ the
flux quantum%
\footnote{From the circuit point of view, these properties make the Josephson
element to behave as a non-linear inductor. Together with the junction
intrinsic capacitance it thus form an anharmonic resonator. The oscillation
mode is known as the {}``plasma resonance''.%
}. Since its prediction, the Josephson current has been observed in
a great variety of systems involving superconducting electrodes connected
by a “weak link”. A weak link can be an insulating layer, as originally
proposed by Josephson, or any quantum coherent conductor, such as
a short normal (diffusive or ballistic) metallic wire, a point contact,
a carbon nanotube, or a graphene layer \citep{golubov_current-phase_2004}.
A great deal of theoretical activity has been devoted to relate the
maximum (or critical) supercurrent $I_{0}$ of a weak link to its
normal state properties. A unifying theoretical description providing
this relation for an arbitrary structure in terms of its transmission
PIN emerged in the 1990s and is based on the concept of Andreev bound
states \citep{furusaki_unified_1990,beenakker_josephson_1991}.

Due to the existence of a gap in the quasiparticle density of states
in a superconductor, an electron impinging from a normal metal into
a superconducting electrode with an energy $|E|<\Delta$ cannot enter
it as such, and should in principle be reflected back. However, if
there is no barrier at the interface it will be {}``Andreev reflected''
as a hole at the same energy and moving in the opposite direction,
adding a Cooper pair to the superconductor \citep{andreev__1964}.

\subsection{Andreev Bound States}

\begin{figure}[tbh]
\begin{centering}
\includegraphics[height=120pt]{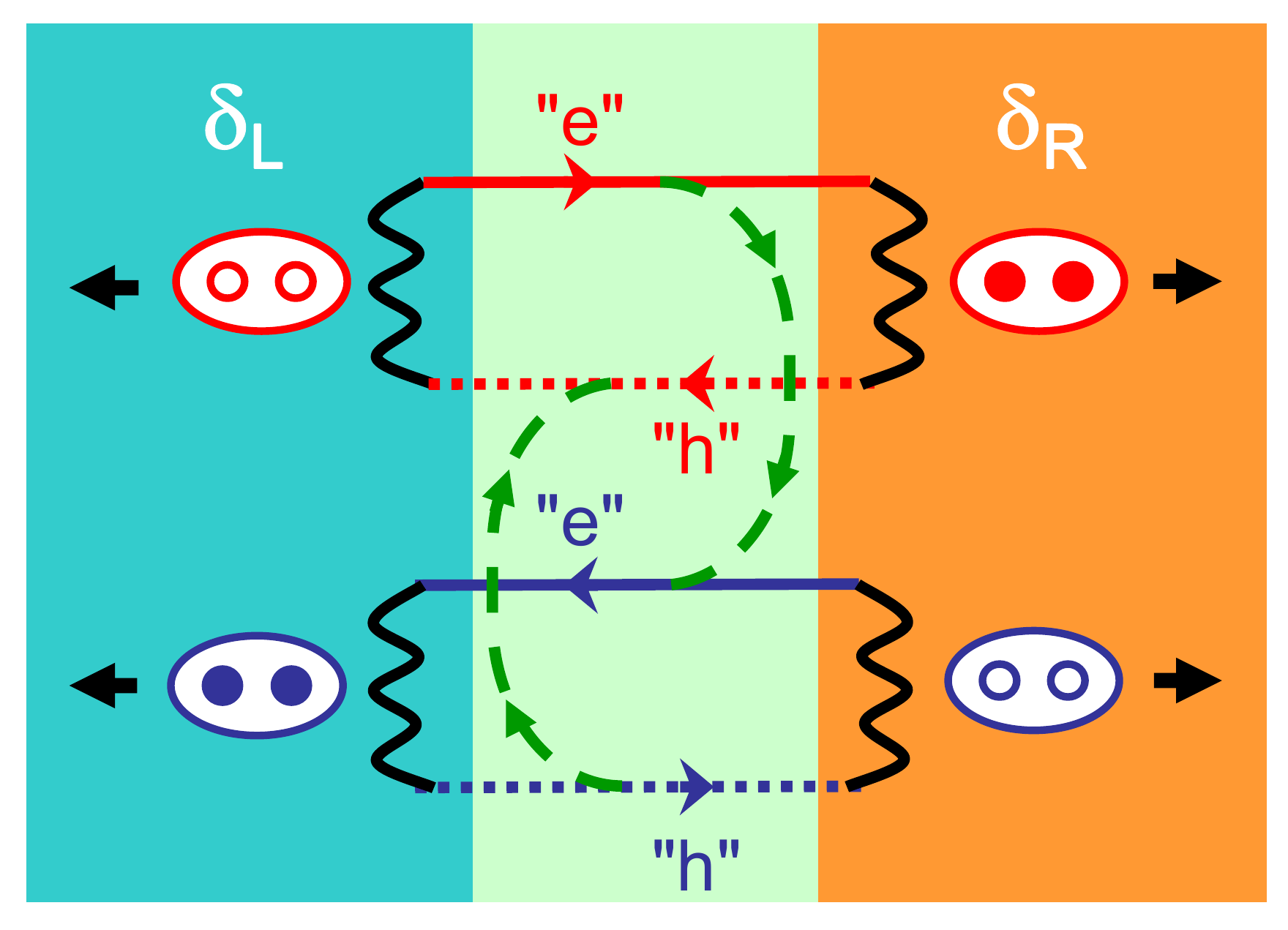}
\par\end{centering}

\caption{Schematic representation of the microscopic processes leading to the
formation of two Andreev bound states in a short channel connecting
two superconducting electrodes with different phases $\delta_{L}$
and $\delta_{R}$. The wiggly lines represent an Andreev reflection
in which an electron (hole) is reflected as a hole (electron) acquiring
the local superconducting phase. The upper (lower) loop corresponds
to the transfer of Cooper pairs to the right (left). When the transmission
probability of the channel is below unity, in addition to the Andreev
reflection processes, normal reflection processes (dashed lines) connects
electron (hole) states traveling in opposite directions.\label{fig:AndreevLoopBal}}
\end{figure}

The formation of Andreev bound states is illustrated for a single
channel in Fig.~\ref{fig:AndreevLoopBal} \citep{bagwell_suppression_1992}.

In a perfectly transmitting channel, a right-moving electron is Andreev
reflected at the interface with the right superconducting electrode
into a left-moving hole, and acquires an energy dependent phase shift
$\arccos\left(E/\Delta\right)+\delta_{R}$ where $\delta_{R}$ is
the phase of the local superconducting order parameter. In turn, this
left-moving hole is Andreev reflected as a right-moving electron at
the left electrode, removing a Cooper pair from it. These successive
reflections interfere constructively, like in a Fabry-Pérot interferometer,
when the global phase shift acquired along one round-trip is an integer
multiple of $2\pi$. Of course, a similar process occurs for left-moving
electrons reflected as right-moving holes, and eventually two resonant
quasiparticle states $\left|\leftarrow\right\rangle \mathrm{\, and}\,\left|\rightarrow\right\rangle $
appear in the channel region. Their phase-dependent energies lie within
the superconducting energy gap and are symmetric with respect to the
Fermi level.

\begin{figure}[tbh]
\begin{centering}
\includegraphics[height=120pt]{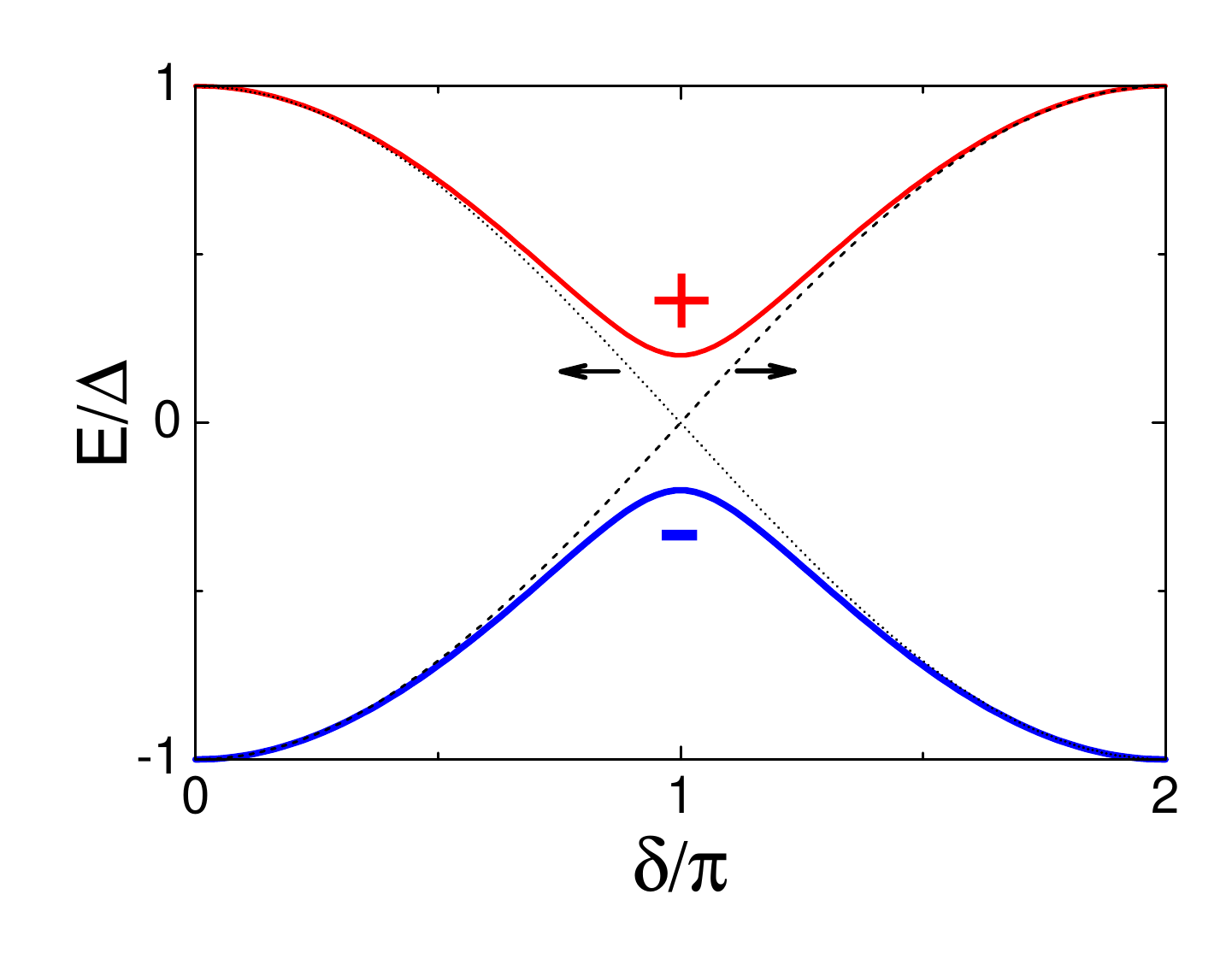} \hspace{2cm}\includegraphics[height=120pt]{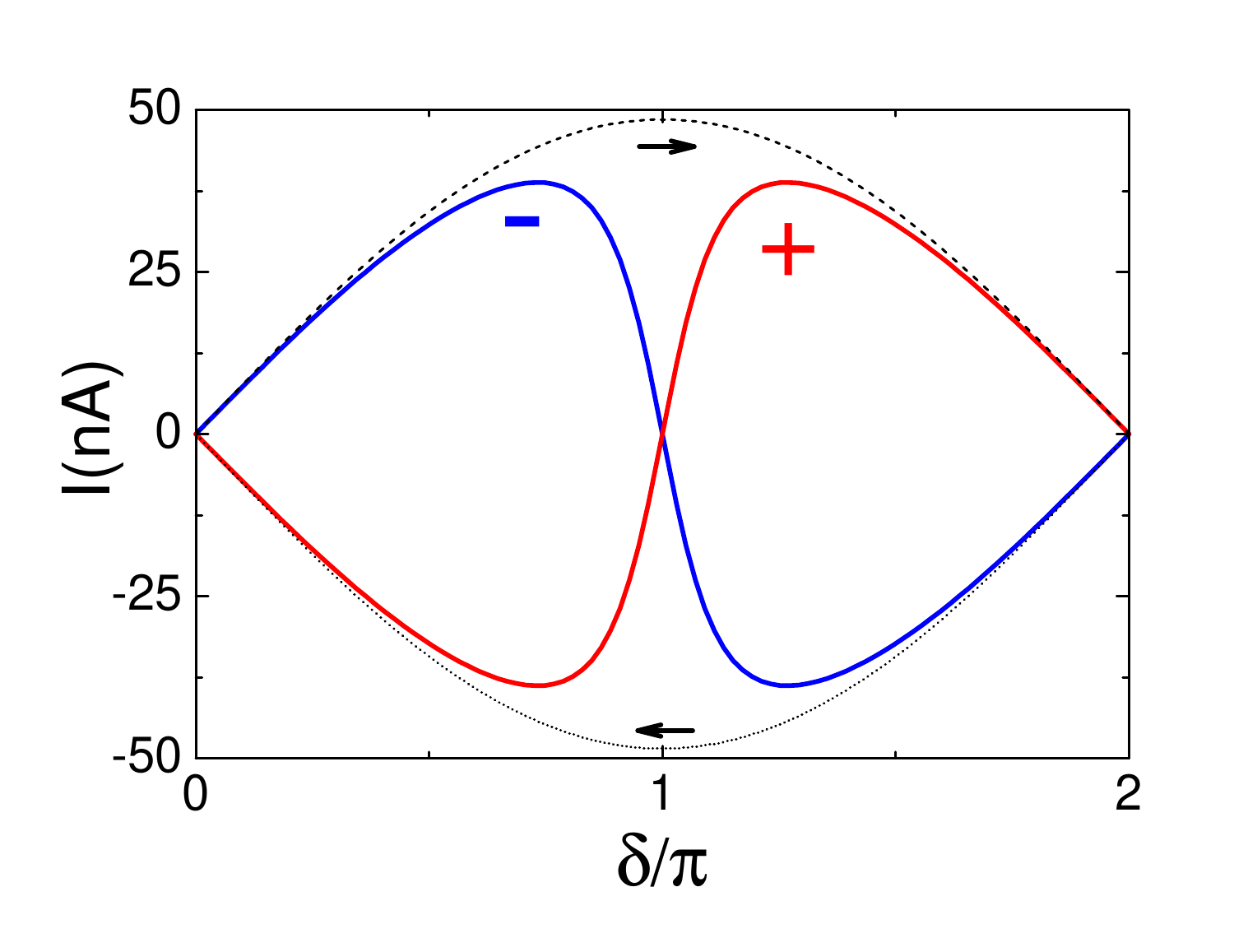}
\par\end{centering}

\caption{\textbf{Left:} Energy of Andreev states $\left|-\right\rangle \mathrm{\, and}\,\left|+\right\rangle $
in a single channel of transmission $\tau=0.96$ (solid curves) as
a function of the superconducting phase difference $\delta$ across
it. The dashed curves represent those in a reflectionless channel
$\left|\leftarrow\right\rangle \mathrm{\, and}\,\left|\rightarrow\right\rangle $.
\textbf{Right:} Supercurrent carried by the upper $\left|+\right\rangle $
and lower $\left|-\right\rangle $ Andreev states calculated for $\Delta=200\:\mathrm{\mu eV}$
(full lines). The dashed curves are the currents for the reflectionless
channel.\label{fig:Andreev}}
\end{figure}
If the channel is not perfectly transmitting ($\tau<1$), a right-moving
electron can also be simply reflected as a left-moving electron. The
existence of a finite reflection probability thus couples the two
reflectionless states. The two resulting states, denoted $\left\{ \left|-\right\rangle ,\left|+\right\rangle \right\} $,
are called the {}``Andreev states'', and have energies $\mp E_{A}$
shown in Fig.~\ref{fig:Andreev}, where: \begin{equation}
E_{A}\left(\delta,\tau\right)=\Delta\sqrt{1-\tau\sin^{2}\left(\frac{\delta}{2}\right)}.\end{equation}
Let us point out that in the tunnel limit $\tau\rightarrow0,$ this
expression correctly yields the $\cos\left(\delta\right)$ dependence
of the coupling energy predicted by Josephson for a tunnel junction.
The transition frequency \textcolor{black}{between the two Andreev
states}\textcolor{red}{{} }reaches its minimum $\left(2\Delta/h\right)\sqrt{1-\tau}$
at $\delta=\pi$.

Because the number of transferred pairs and the phase difference are
conjugated variables, these states carry supercurrents given by: \begin{equation}
I_{\pm}\left(\delta,\tau\right)=\frac{2\pi}{\phi_{0}}\frac{\partial E_{\pm}\left(\delta\right)}{\partial\delta}=\mp\frac{e\Delta}{2\hbar}\frac{\tau\sin\delta}{\sqrt{1-\tau\sin^{2}\left(\frac{\delta}{2}\right)}}.\label{eq:CurrPhase}\end{equation}
and shown in the right panel of Fig.~\ref{fig:Andreev}. Since at
a given phase the two Andreev bound states carry the same current
but in opposite directions, the net supercurrent results from the
imbalance of their populations. At zero temperature, only the lower
energy state is occupied and the maximum (or critical) supercurrent
is given by\[
I_{0}\left(\tau\right)=\frac{e\Delta}{\hbar}\left(1-\sqrt{1-\tau}\right),\]
which is not simply proportional to $\tau$ and thus neither to the
normal state conductance. In aluminum, where the superconducting gap
$\Delta$ is typically of the order of $200\:\mathrm{\mu eV}$, this
current reaches at most $\simeq50\:\mathrm{nA}$ for a single perfectly
transmitting channel, and the transition frequency is in the microwave
range.%
\begin{figure}[tbh]
\begin{centering}
\includegraphics[height=125pt]{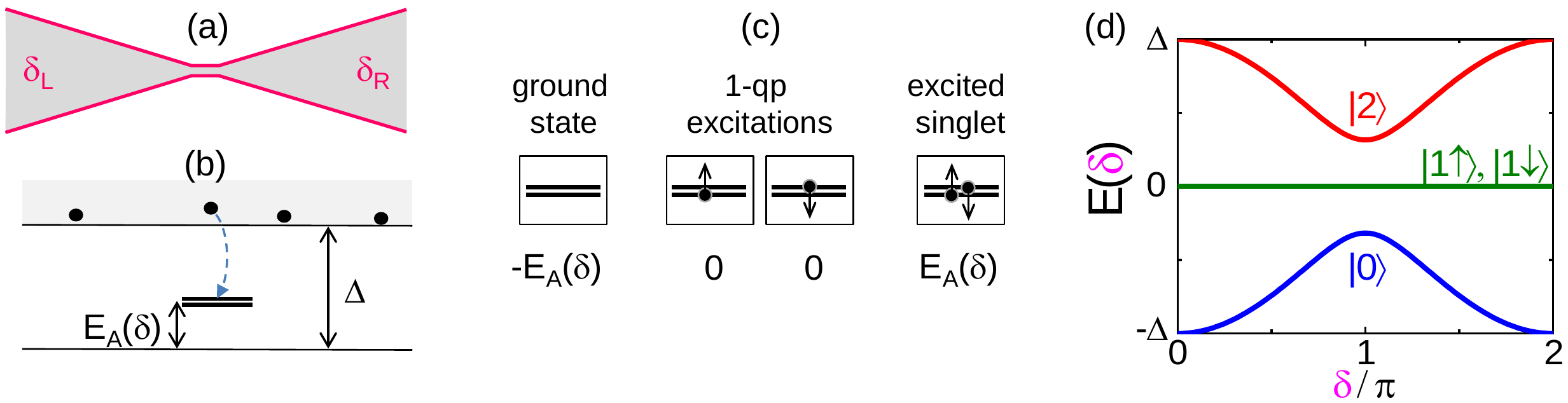}
\par\end{centering}

\caption{\textbf{\emph{(a)}} Short one-channel constriction between two superconducting
electrodes (phase difference $\delta=\delta_{L}-\delta_{R}$).\emph{
}\textbf{\emph{(b)}}\emph{ }Excitation spectrum: besides the usual
continuum of excited states above the energy gap $\Delta$, that extends
all across the structure, there is at the constriction a discrete
Andreev spin-degenerate doublet with energy $E_{A}(\delta)$ above
the ground state, where quasiparticles can get trapped. \textbf{\emph{(c)}}
Four possible configurations of the Andreev doublet at the constriction:
in the global ground configuration, labeled $\left|0\right\rangle $,
the doublet is empty (it corresponds to just the ground Andreev state
$\left|-\right\rangle $ of Fig.~\ref{fig:Andreev} being occupied);
in the two odd configurations, each with definite spin $\pm1/2$ and
labeled $\left|1\uparrow\right\rangle $ and $\left|1\downarrow\right\rangle $,
one quasiparticle is added to the contact (they correspond to the
Andreev states of Fig.~\ref{fig:Andreev} being both occupied or
both empty) \citep{chtchelkatchev_andreev_2003}; the last configuration
is a spin-singlet double excitation, labeled $\left|2\right\rangle $
(it corresponds to just the excited Andreev state $\left|+\right\rangle $
of Fig.~\ref{fig:Andreev} being occupied). \textbf{\emph{(d) }}Total
energy of the four configurations vs phase . The energy of the ground
configuration $\left|0\right\rangle $ is that of the lowest Andreev
level $-E_{A}$. The two odd configurations, $\left|1\uparrow\right\rangle $
and $\left|1\downarrow\right\rangle $, have zero energy. The double
excitation configuration $\left|2\right\rangle $, has the energy
of the excited Andreev state $+E_{A}$.}
\label{Fig: excitation spectrum}

\end{figure}

\subsection{Excitation spectrum}

It is important to note that the pair of Andreev bound states leads
locally to an excitation spectrum (Fig.\,\ref{Fig: excitation spectrum})
containing, at an energy $E_{A}\leq\Delta$ above the ground state,
a discrete spin-degenerate doublet \citep{chtchelkatchev_andreev_2003}.
There are four possible occupation configurations for the system.
In the global ground state configuration, the total energy contains
a phase dependent term $-E_{A}(\delta,\tau)$ arising from the lowest
lying localized Andreev state, which is occupied by a spin-singlet
pair of electrons and gives rise to a supercurrent $I_{-}=-\varphi_{0}^{-1}\partial E_{A}/\partial\delta,$
with $\varphi_{0}=\hbar/2e.$ Then, there are two {}``odd'' configurations
(spin 1/2) with a single excitation of the doublet at $E_{A}$,\emph{
}corresponding to a quasiparticle trapped in the one-channel constriction.
In this case the global energy is zero,\emph{ }phase independent,
and the total supercurrent is zero. Finally, there is another spin-singlet
configuration with a double excitation which carries a supercurrent
$I_{+}=+\varphi_{0}^{-1}\partial E_{A}/\partial\delta$ exactly opposite
to the one in the ground configuration. Hence the supercurrent through
the constriction is a probe of the excitation configuration of the
system.

\subsection{Voltage bias and Andreev states}

When the structure is voltage-biased at $V\ll\Delta/e$, the phase
varies linearly with time at a speed $\dot{\delta}=2\pi V/\phi_{0}$.
When the latter is low enough, one can assume that the Andreev levels
move adiabatically within the superconducting gap $\Delta$. As the
motion is periodic, there is no energy transfer to the system on average
and a purely ac current flows. This corresponds to the second striking
prediction of Josephson, the ac Josephson effect. For larger voltages,
and therefore larger speeds, non-adiabatic transitions (Landau-Zener
type) arise between the Andreev levels, giving rise to the onset of
the dissipative MAR current discussed in section 2.2 \citep{averin_ac_1995}.
The onset of this dissipative current depends on the channel transmission.

\section{Probing the Josephson effect in atomic contacts}

For an arbitrary quantum coherent conductor the phase-driven supercurrent
is governed by the occupation numbers of the Andreev states in all
channels. At very low temperature, when only the lower energy Andreev
state of each channel is occupied, the current-phase relation is:
\begin{equation}
I\left(\delta,\left\{ \tau_{i}\right\} \right)=\frac{e\Delta}{2\hbar}\sum_{i=1}^{N}\frac{\tau_{i}\sin\left(\delta\right)}{\sqrt{1-\tau_{i}\sin^{2}\left(\frac{\delta}{2}\right)}}.\label{eq:Curr-phase N}\end{equation}
This is one of the fundamental predictions of the mesoscopic theory
of the Josephson effect that we have probed in our experiments with
atomic contacts.

In a first series of experiments carried out on current-biased contacts
\citep{goffman_supercurrent_2000}, we measured the average supercurrent
$I_{S}$ at which the system switches to the dissipative branch. In
this case, the phase is not an externally tunable parameter, but a
dynamical variable moving in a potential landscape fixed by the total
Josephson coupling energy and submitted to random fluctuations imposed
by the biasing circuit. As a consequence the switching current is
always smaller than the critical current $I_{0}.$ However, because
we embedded the contact in a properly designed on-chip circuit, dissipation
and thus fluctuations were under control. Most of the experimental
results could be well understood by considering just the contribution
of the ground Andreev state of each channel, \textcolor{black}{but
evidence for non-adiabatic transitions between Andreev states was
observed for nearly ballistic contacts.}

However, these experiments were actually only an indirect test of
the prediction of Eq.~(\ref{eq:Curr-phase N}), since the phase was
not an external parameter that could be swept over its entire range.
A thorough test of the theory actually requires measuring, for a given
atomic contact, both its current-phase relation and its $I(V)$ characteristic
(to determine its PIN code). These two measurements require contradictory
biasing conditions. For the former, the atomic contact must be phase-biased,
which requires inserting it in a small superconducting loop threaded
by a magnetic flux. For the latter, one needs to voltage-bias the
same atomic contact, which cannot be achieved if it is shunted by
the superconducting loop. Obviously, one needs a superconducting reversible
switch in order to toggle between the two biasing conditions without
disturbing the atomic contact.

\subsection{Measurement of the current-phase relation}

\begin{figure}[tbh]
\begin{centering}
\includegraphics[height=140pt]{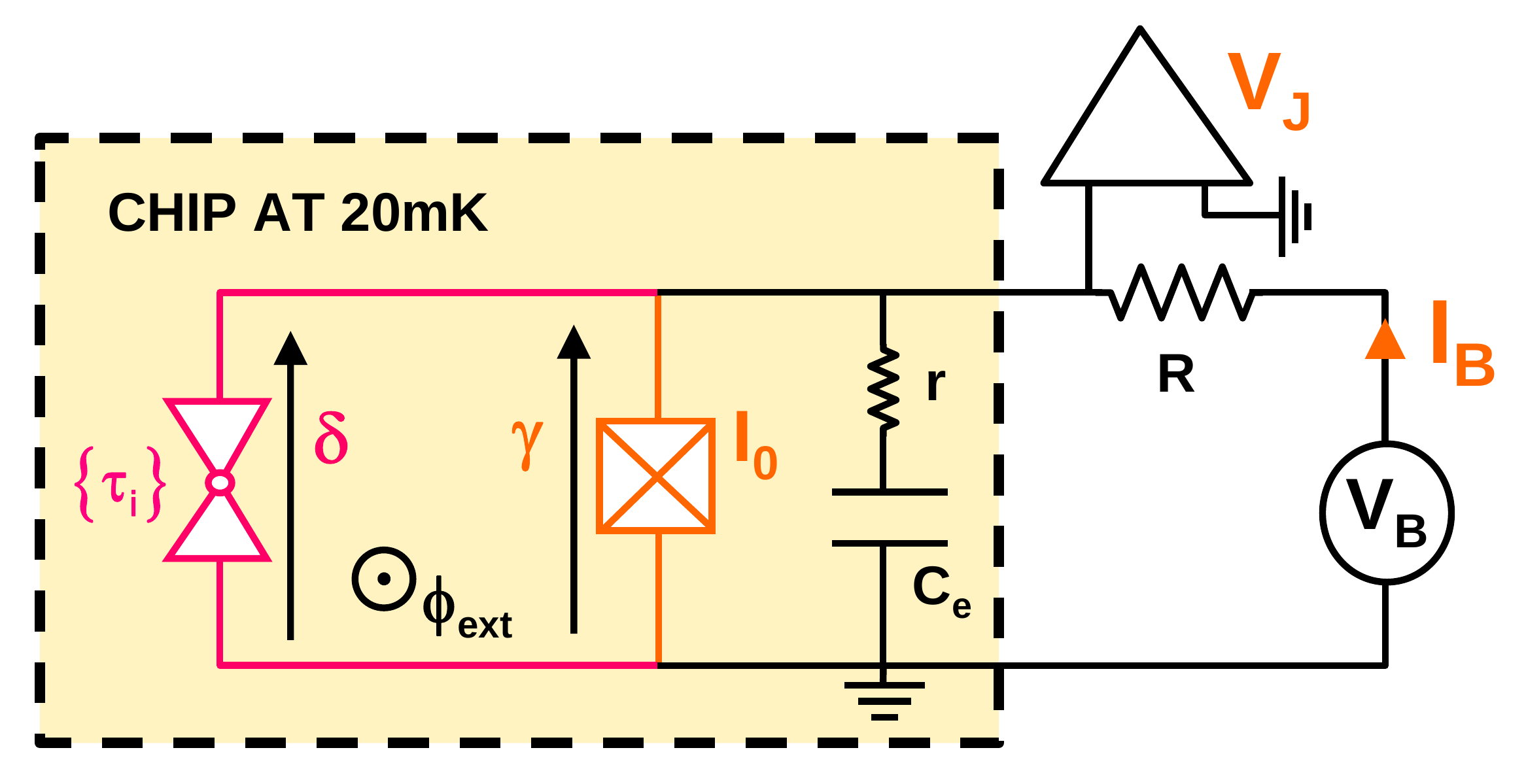}\includegraphics[height=140pt]{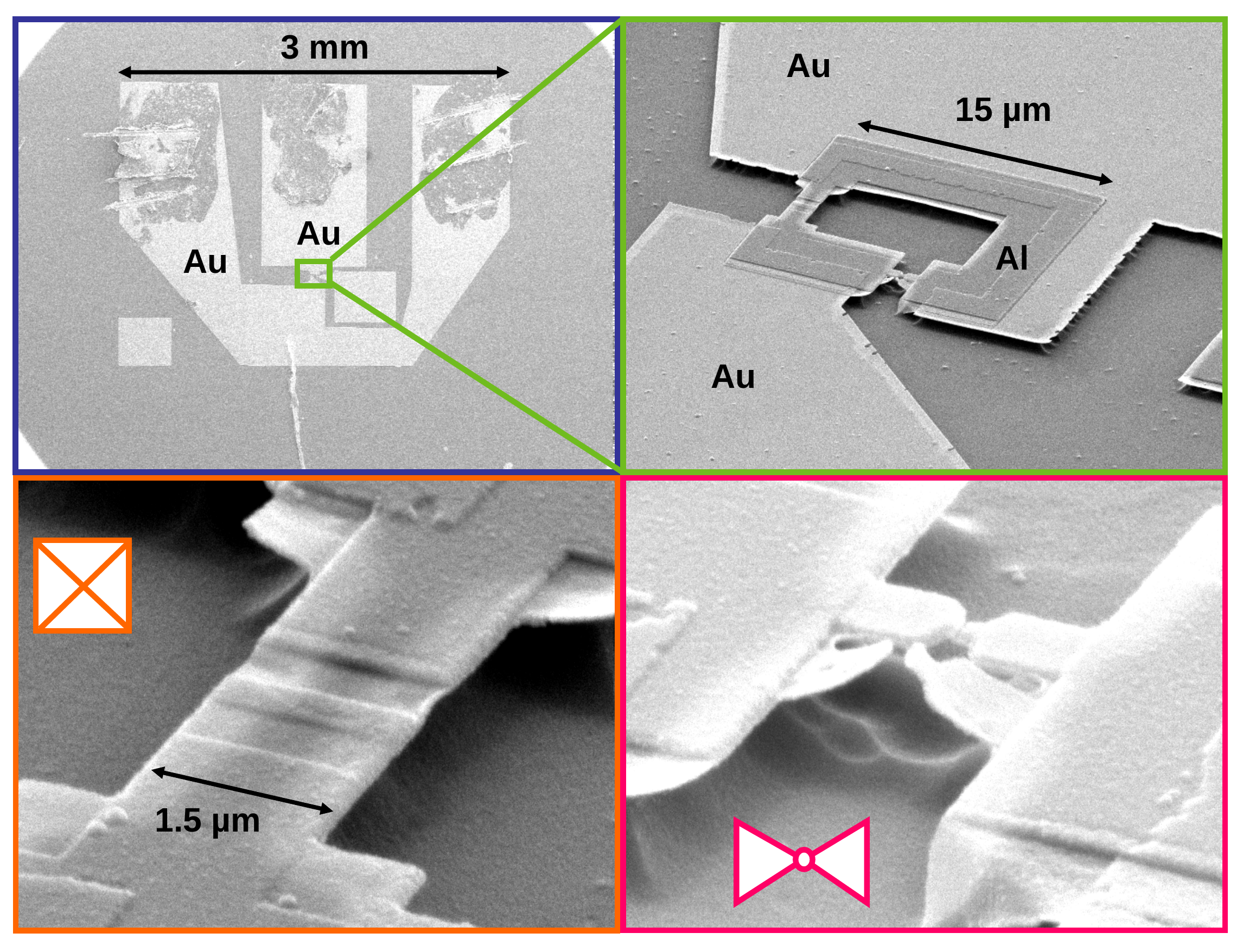}
\par\end{centering}

\caption{\textbf{Left:} An aluminum atomic contact in parallel with a Josephson
junction having a large critical current $I_{0}=310\:\mathrm{nA}$
forms an atomic-SQUID. An on-chip capacitor $C_{e}\simeq20\:\mathrm{pF}$
lowers the plasma frequency of the junction to $1.1\:\mathrm{GHz}$
and the associated resistance $r\sim0.5\mathrm{\Omega}$ damps the
dynamics of the phase. The SQUID is biased through a resistor $R=200\:\mathrm{\Omega}$
by a current $I_{\mathrm{B}}$, and the voltage $V_{\mathrm{J}}$
across it is monitored to detect switching. \textbf{Right:} SEM micrographs
of sample \#1 at different scales. In the upper-right panel, the brighter
pads are gold electrodes, while the darker part constituting the loop
is made out of aluminum. The whole structure is deposited on top of
a polyimide layer on a metallic substrate. The gold electrodes form
the shunting capacitor $C_{\mathrm{e}}$ through the metallic substrate,
and also act as quasiparticle traps (see \citep{huard_interactions_2006,chauvin_effet_2005}
for details). The junction (lower-left panel) is fabricated using
the technique of double-angle evaporation through a suspended mask.
This also leads on the suspended bridge side to an extraneous metallic
bridge, as seen on the bottom-right panel.\label{fig:SetupAC1}}
\end{figure}

In the setup presented in Fig.~\ref{fig:SetupAC1} a Josephson junction
acts as such a switch \citep{della_rocca_measurement_2007,huard_interactions_2006,chauvin_effet_2005}.
The atomic contact and the Josephson junction are embedded in a small
superconducting loop, forming a device that was coined the {}``atomic-SQUID''.
This ancillary junction not only allows both biasing configurations
but is also used to measure the loop current. The critical current
of the Josephson junction is chosen to be much larger than the one
of a typical aluminum one-atom contact ($\lesssim50\:\mathrm{nA}$)
so that the SQUID essentially behaves like a slightly perturbed Josephson
junction.

The $I(V)$ characteristic of the atomic contact, denoted $I_{\textrm{AC}}(V)$,
is obtained from that of the SQUID $I_{\textrm{AS}}(V)$. In principle,
in the region below the superconducting gap $|eV|\leq2\Delta$, the
DC current flowing through the junction is expected to be zero. In
practice however, a sizable current is observed experimentally in
this region in the characteristic $I_{\textrm{JJ}}(V)$ of the junction
alone, which can be measured when the metallic bridge forming the
atomic contact is fully open. Assuming that $I_{\mathrm{JJ}}(V)$
is not affected by the contact, $I_{\textrm{AC}}(V)$ is then obtained
by the subtraction:\begin{equation}
I_{\textrm{AC}}(V)=I_{\textrm{AS}}(V)-I_{\textrm{JJ}}(V),\label{eq:Substract}\end{equation}
 which is then fitted using the MAR theory to obtain the transmissions.

The superconducting loop allows imposing a phase difference across
the atomic contact by applying an external magnetic flux $\phi_{\textrm{ext}}$.
If the loop is sufficiently small so that the screening flux can be
neglected%
\footnote{The geometric inductance $L_{L}$ of the loop, typically $10\:\mathrm{pH}$
is chosen to be negligible as compared to the inductance of both the
Josephson junction $L_{J}\simeq1\:\mathrm{nH}$ and the atomic contact
$L_{\textrm{AC}}\simeq10\:\mathrm{nH}$. In this way, the phase-drop
takes place essentially across these two last elements.%
}, the phase differences $\gamma$ (across the tunnel junction) and
$\delta$ (across the atomic-size contact) are linked through:\begin{equation}
\delta=\gamma+2\pi\frac{\phi_{\textrm{ext}}}{\phi_{0}}\equiv\gamma+\varphi,\label{eq:LinkPhase}\end{equation}
 where $\varphi$ is the reduced flux threading the loop \citep{lefevre-seguin_thermal_1992}.

At zero temperature, the large Josephson junction would switch out
of its zero-voltage state at a phase difference $\gamma=\frac{\pi}{2}$
, when it carries exactly its critical current $I_{0}$. Assuming
that the contributions of the junction and of the atomic contact can
be separated, the critical current $I_{\textrm{AS}}^{0}$ of the SQUID
is the sum of the critical current $I_{0}$ of the junction and of
the flux-dependent critical current $I_{\textrm{AC}}(\delta)$ of
the atomic contact:\begin{equation}
I_{\textrm{AS}}^{0}(\varphi)=I_{0}+I_{\textrm{AC}}(\frac{\pi}{2}+\varphi).\label{eq:ModCurveIdeal}\end{equation}
 Measuring the flux dependence of the critical current of the SQUID
would thus be a direct way to probe the current-phase relation of
the atomic contact. However, as mentioned before, in practice the
critical current of a small Josephson device is not accessible. Due
to fluctuations the system switches stochasticaly to a dissipative
state before reaching this maximum current, and one simply measures
the mean switching current of the SQUID.

\begin{figure}[tbh]
\begin{centering}
(a)\includegraphics[height=150pt]{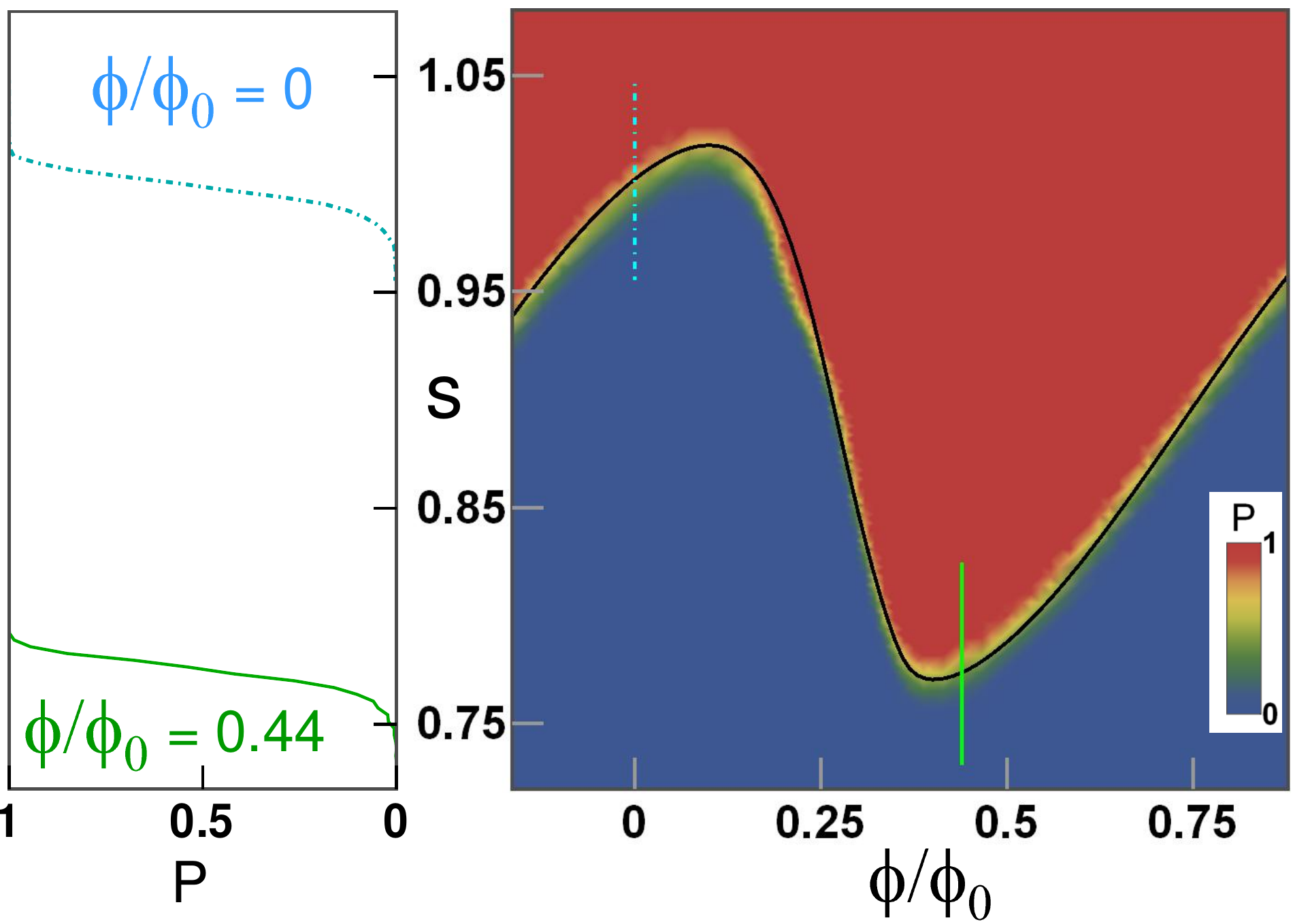}\hspace{1cm}
(b)\includegraphics[height=150pt]{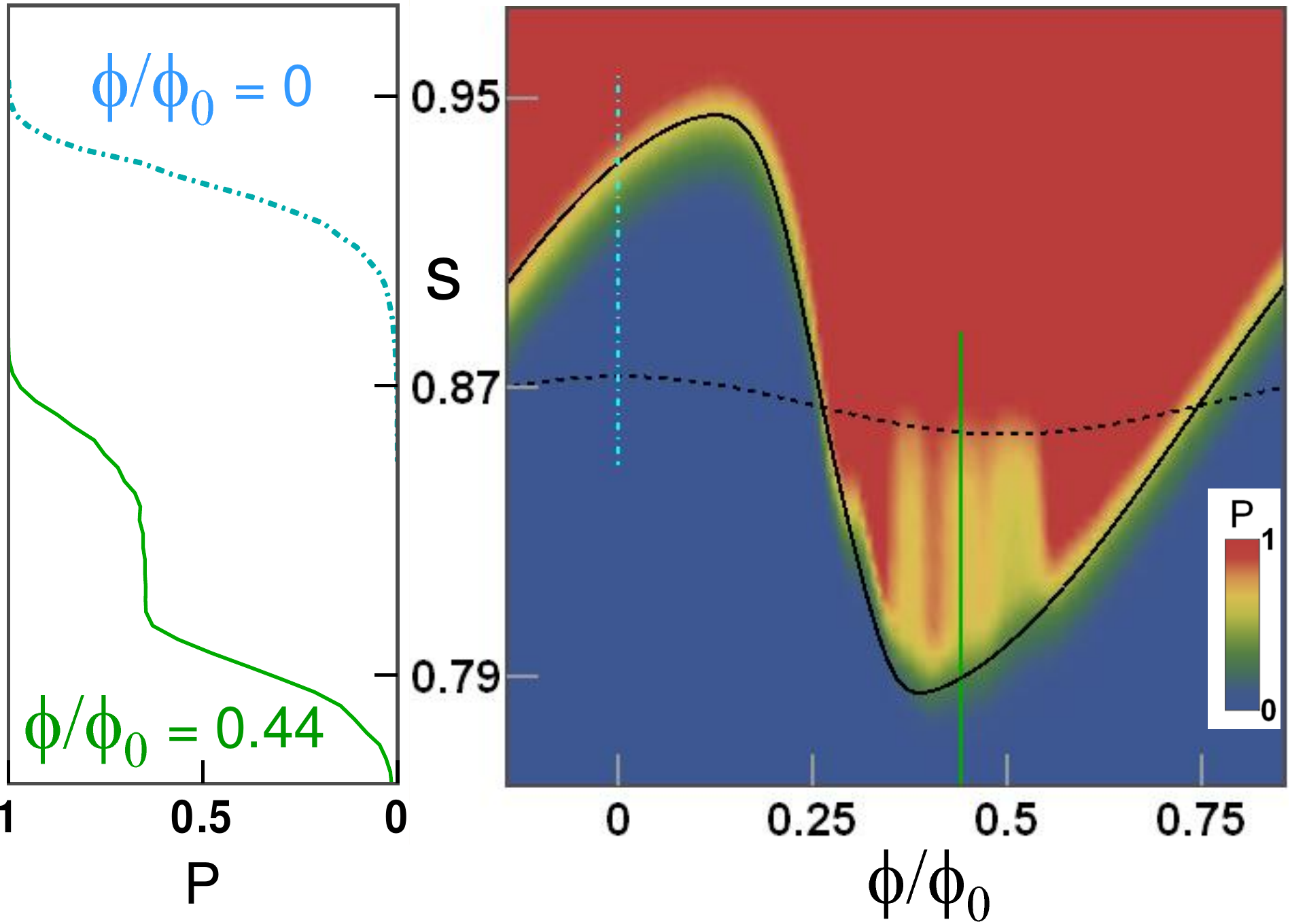}
\par\end{centering}

\caption{\textbf{(a) Left:} Switching probability $P$ for an atomic SQUID
obtained with sample \textbf{\#1} as a function of the reduced bias
current $s$, for two values of the external flux. \textbf{Right:}
switching probability (color-coded) as a function of both the reduced
bias current and flux. The two vertical lines correspond to the two
curves plotted in the left panel. The full black line corresponds
to the modulation curve $s^{*}(\varphi)$ calculated using the full
PIN of the contact, taking into account the contribution of just the
Andreev ground state in each channel. The PIN code of the atomic contact
was $\left\{ 0.969,\:0.137\right\} .$\textbf{}\protect \\
\textbf{(b) Left:} switching probability $P$ for an atomic-SQUID
obtained with sample\textbf{ \#2} as a function of the reduced bias
current $s$, for two values of the external flux. The upper curve
is similar to those obtained in sample \textbf{\#1} but the lower
curve presents a clear intermediate plateau. The PIN code of the atomic
contact was $\left\{ 0.994,\:0.097,\:0.096\right\} .$\textbf{ Right:}
switching probability (color-coded) as a function of both the reduced
bias current and flux. The two vertical lines correspond to the two
curves plotted in the left panel. The full line corresponds to the
modulation curve $s^{*}(\varphi)$ calculated taking into account
the contribution of the Andreev ground state of all the channels.
The dashed line corresponds to the modulation curve calculated for
$\left\{ 0.097,\:0.096\right\} $, which excludes the contribution
of the most transmitted channel.\label{fig:modPswit}}
\end{figure}
The mean switching current is determined accurately from the switching
probability $P$. This is measured using a train of bias-current pulses
of a given height and counting the number of pulses for which a voltage
appears across the SQUID. We plot in the right panel of Fig.~\ref{fig:modPswit}(a)
the switching probability as a function of both the reduced bias current
$s=I_{B}/I_{0}$ and the reduced flux $\varphi$ in the loop, for
a particular atomic-contact obtained with sample \#1\textcolor{black}{(described
in Fig.~\ref{fig:SetupAC1}).} The left panel of the same figure
shows two examples of $P(s)$ for two different values of flux. One
observes that the switching probability evolves very rapidly from
0 to 1 in a narrow range of bias current which depends on the applied
flux. In the following, we denote $s^{*}$ the reduced bias current
leading to an switching probability $P=0.5$, and refer to $s^{*}(\varphi)$
as the SQUID {}``modulation curve''.

In order to calculate the expected modulation curve one considers
the phase dynamics in the full Josephson potential of the SQUID in
the presence of thermal fluctuations \citep{della_rocca_measurement_2007,le_masne_asymmetric_2009}.
The potential is dominated by the Josephson energy of the junction
but contains a contribution which consists of the sum of the ground
sate Andreev energies of every channel of the atomic contact. As shown
in the right panel of Fig.~\ref{fig:modPswit}(a) this model works
very well.

In conclusion for all the contacts that we investigated using sample
\#1 the current-phase relation measured at low temperatures is in
quantitative agreement with the mesoscopic description of the Josephson
effect, considering that only the ground Andreev state is occupied
in each channel. The next step in order to achieve a comprehensive
test of the theory would be to probe the excited Andreev states for
example by directly performing spectroscopy of the transitions between
Andreev levels. However, during our first attempts to do so, we observed
unexpected but important phenomena which shed light on the microscopic
nature of the system.

\subsection{Out-of-equilibrium effects}

The experiments presented in the previous sections were performed
with atomic-SQUIDs obtained on sample \#1, which was fabricated on
top of a metallic substrate and in which the on-chip lines connecting
the SQUID to the outside world were made out of gold, a dissipative
metal at low temperature. On the contrary in our first design to attempt
microwave Andreev spectroscopy, in order to minimize dissipation which
is expected to limit the life-time of the excited Andreev levels \citep{desposito_controlled_2001}
and thus broaden the spectral lines, we decided to fabricate the samples
on Kapton substrates with all on-chip electrodes made out of aluminum.
Apart from this change of material sample \#2 was almost identical
to the previous one and in particular the critical current of the
junction was $I_{0}=295\:\mathrm{nA}$, less than $5\%$ smaller than
before. However, as we will discuss now, this change of material had
strong consequences on the overall behavior of the system.

\begin{figure}[tbh]
\begin{centering}
\includegraphics[height=160pt]{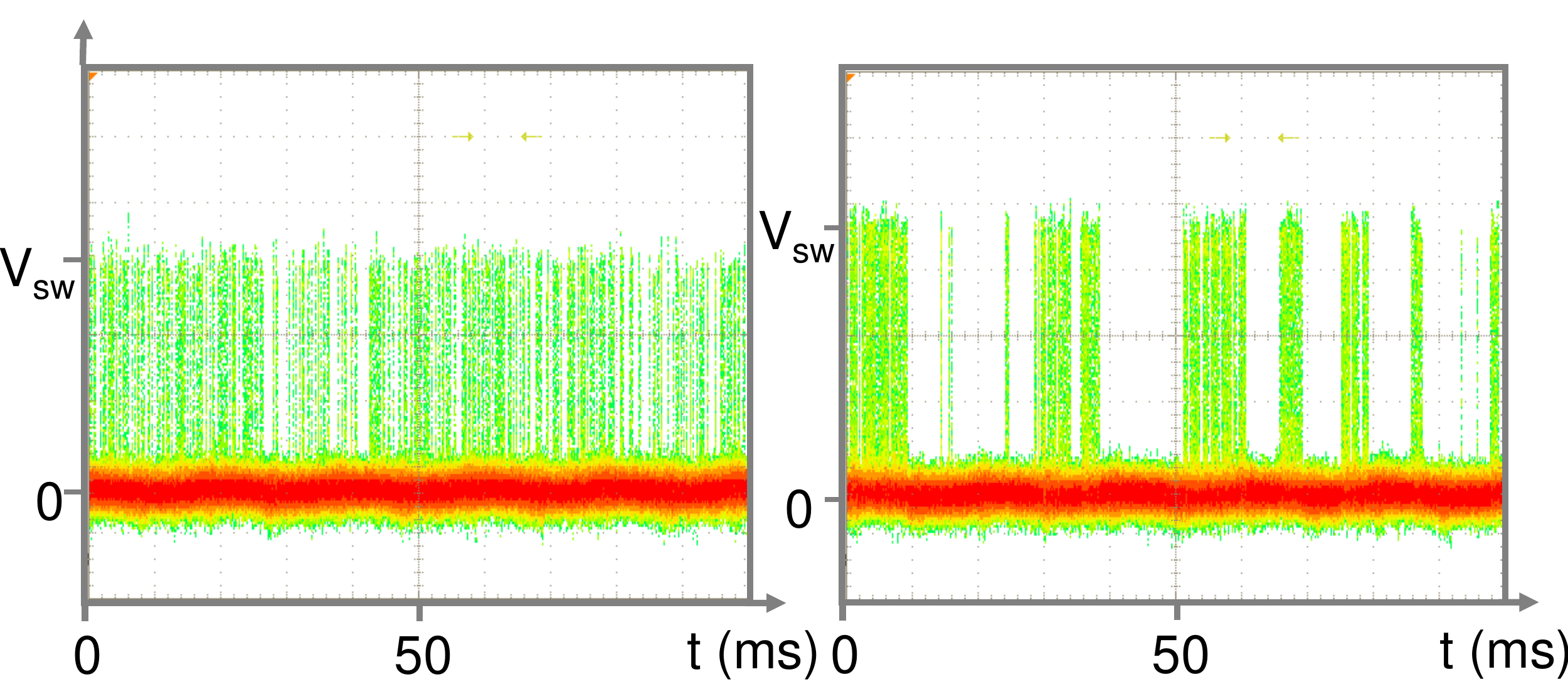}
\par\end{centering}

\caption{Voltage across the atomic-SQUID as a function of time during a sequence
of bias-current pulses ($50\:\mu\mathrm{s}$ period) \textcolor{black}{corresponding
to a switching probability $P=0.2$.} Each voltage pulse corresponds
to a switching event. \textbf{Left}: at a flux value for which the
$P\left(s\right)$ curve is normal. \textbf{Right}: Similar measurement,
at a flux value for which $P\left(s\right)$ displays a step, showing
intermittency attributed to the dynamics of single quasiparticle trapping
and untrapping in the Andreev doublet.\label{fig:Bunching}}
\end{figure}
The right panel of Fig.~\ref{fig:modPswit}(b) shows the measured
switching probability as a function of both the flux and the reduced
bias current for a SQUID with an atomic contact of transmissions $\left\{ 0.994,\:0.097,\:0.096\right\} $in
sample\#2. As compared to the previous sample (see Fig.~\ref{fig:modPswit}(a))
the switching probability $P\left(s\right)$ does not vary sharply
from 0 to 1 for all values of the flux. In some flux ranges, it displays
a step as a function of the bias current, as shown for $\varphi=0.44$
in the left panel of the same figure. These features appear only in
a region bounded by the expected modulation curve $s^{*}(\varphi)$
and by a second modulation curve calculated without the contribution
of the highest transmitted channel. In other words, in this region
the highest transmitted channel does not seem to always contribute
to the total current through the contact. This is exactly what is
expected if the two Andreev states of this channel are either both
empty or both occupied, an occupation configuration achieved when
a single quasiparticle is trapped in the Andreev doublet as discussed
in section 3.2. Since the Andreev states have energies below the superconducting
gap the quasiparticles are easily trapped. The fact that this effect
only appears on samples having no large normal electrodes able to
efficiently remove the quasiparticles strongly supports this hypothesis.
Moreover, if trapping happens in a flux region for which the critical
current gets enhanced by the double occupancy, subsequent current
pulses of the same height would be unable to make the system switch.
It is only after the quasiparticle gets untrapped that switching occurs
again. This results in intermittency of switching as quasiparticles
trap and untrap, as shown in Fig.~\ref{fig:Bunching}. Recently,
we have measured the full dynamics of trapping and found that the
lifetime of trapped quasiparticles can be quite long, on the order
of $100\,\mathrm{\mu s}$ \citep{zgirski_evidence_2011}. \textcolor{black}{This
phenomenon is reminiscent of  the {}``poisoning'' by a single quasiparticle
observed in single Cooper-pair devices containing small superconducting
islands in which the parity of the total number of electrons actually
matters. However, in the case we discuss here quasiparticle trapping
occurs in a constriction between two superconductors, a system containing
no island at all and where charging energy does not play a role.}\textcolor{red}{{} }

\section{Towards Andreev spectroscopy}

As shown in the previous section, up to now only the properties of
the Andreev ground state have been probed in detail. Performing spectroscopy
of the transition between two Andreev states is clearly the next important
step which we are presently pursuing along two different lines.

\subsection{Microwave reflectometry}

\begin{figure}[tbh]
\begin{centering}
\includegraphics[height=120pt]{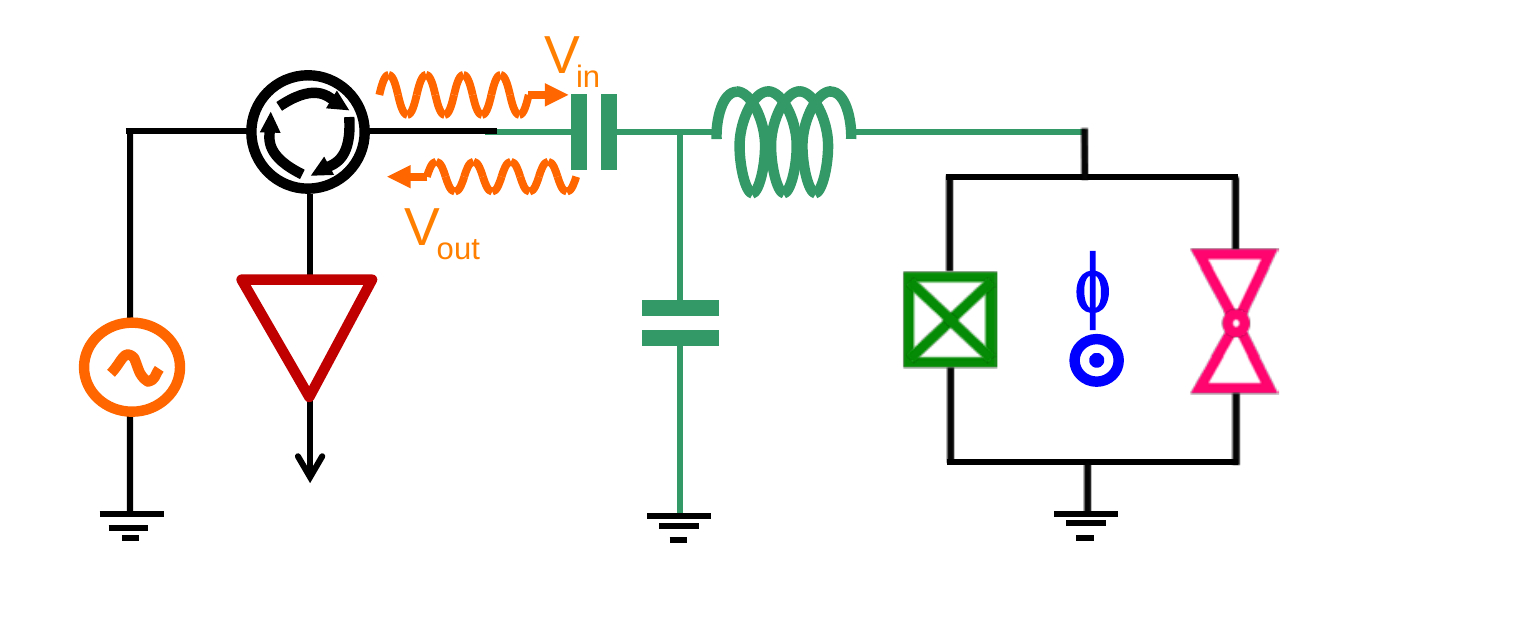}\includegraphics[height=120pt]{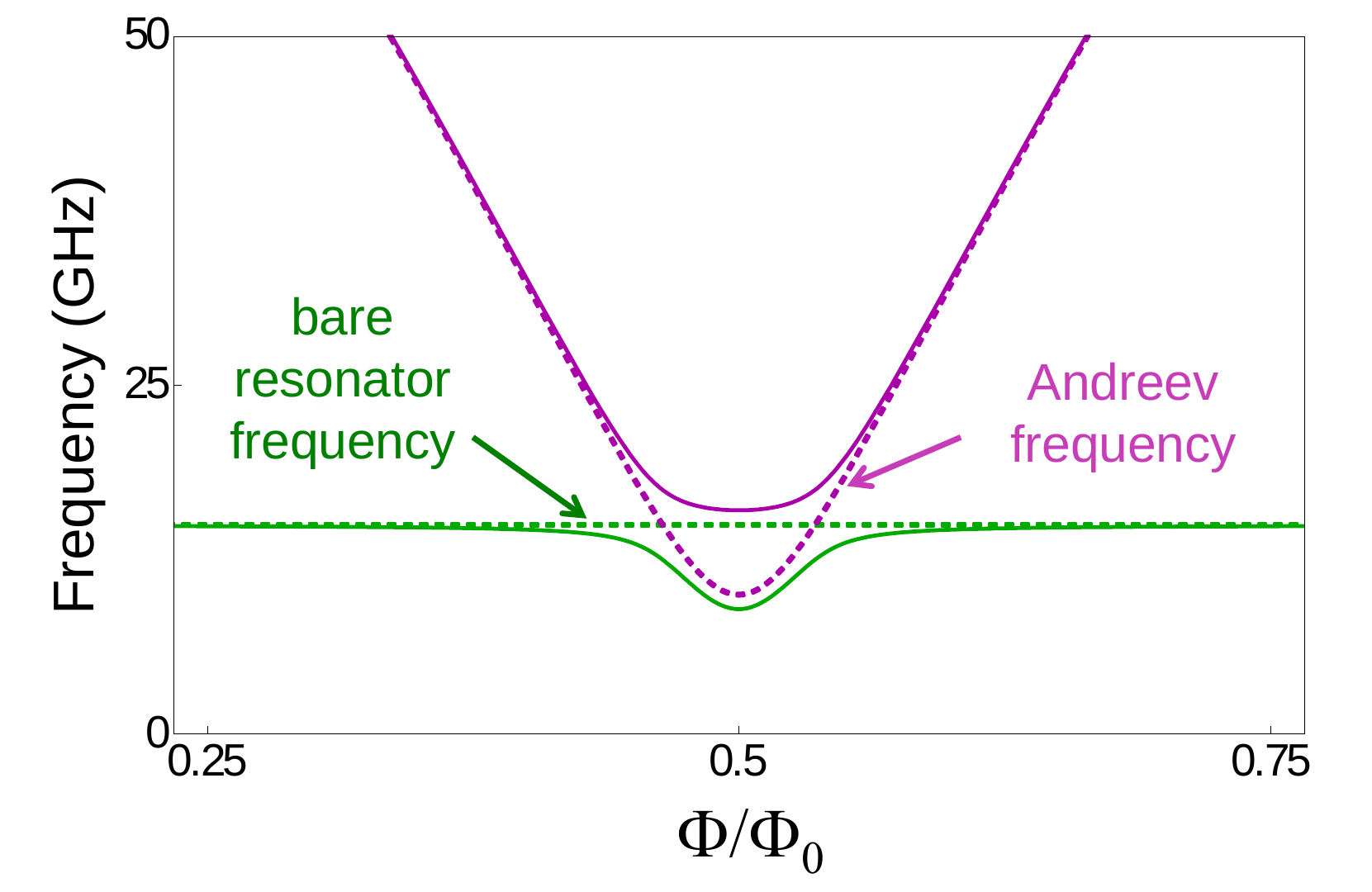}
\par\end{centering}

\caption{\textbf{Left:} An atomic-SQUID coupled to a microwave resonator, represented
here by the LC circuit (in green). The spectrum of the combined system
is probed through microwave reflectometry by weakly coupling the resonator
to the external setup through a small capacitor. The signal reflected
by the sample is diverted by a circulator into a microwave amplifier.
\textbf{Right:} example of the expected spectrum (full lines) as a
function of the magnetic flux threading the SQUID loop. The resonance
frequency of the bare resonator (green dashed line) is chosen to be
$15\,\mathrm{GHz,}$ the Andreev frequency (magenta dashed line) is
for a channel of $\tau=0.995,$ and the coupling energy is $h\times\left(5\,\mathrm{GHz}\right).$
Note the anti-crossing signaling the hybridization of the two quantum
systems.}
\label{Fig: cavity}%
\end{figure}
 The idea here is to couple an atomic SQUID to the electromagnetic
field of a coplanar waveguide resonator (left panel of Fig.~\ref{Fig: cavity}).
By varying the flux threading the SQUID loop the Andreev transition
frequency can be brought into resonance with one of the resonator's
modes. This will result in the hybridization between the Andreev levels
and the modes of the cavity (right panel of Fig.~\ref{Fig: cavity}).
The goal of the experiment would be to detect this hybridization,
a strategy that has been implemented successfully for superconducting
qubits \citep{blais_cavity_2004}. The required technical developments
are underway.

\subsection{Inelastic Cooper pair tunneling}

\begin{figure}[tbh]
\begin{centering}
\includegraphics[height=120pt]{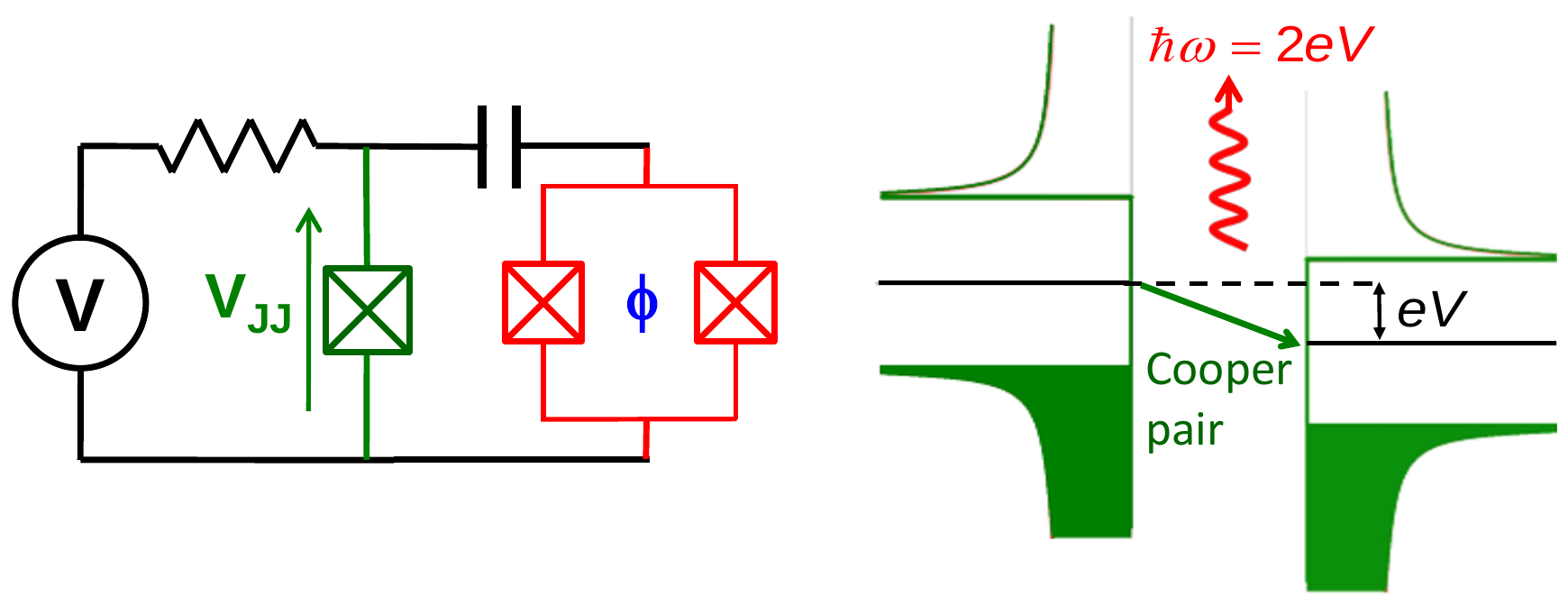}\includegraphics[height=120pt]{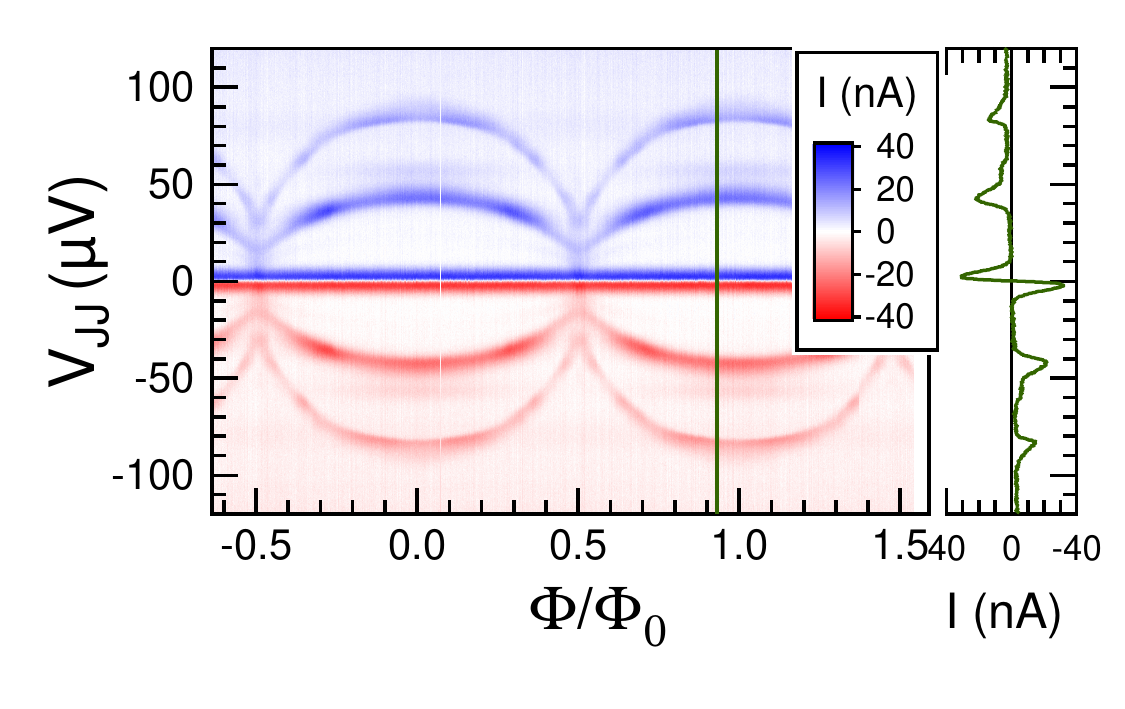}
\par\end{centering}

\caption{\textbf{Left:} Experimental setup: a voltage biased Josephson junction
capacitively coupled to a dc SQUID. \textbf{Center:} Density of states
of the two superconducting electrodes of the junction, shifted by
$eV.$ Cooper pairs can tunnel only if their energy drop $2eV$ is
absorbed as photons in the modes of its electromagnetic environment.
\textbf{Right:} Measured current (color-coded) through the junction
as a function of the voltage $V_{\mathrm{JJ}}$ across it (vertical
axis) and of the magnetic flux threading the SQUID (horizontal axis).
The flux dependent lines correspond to the excitation of the plasma
mode of the SQUID. An example of the junction $IV$ (for a flux $\phi\simeq.92\phi_{0}$)
is shown in the rightmost panel.}
\label{Fig: ICPT}%
\end{figure}
The second strategy that we are implementing to perform spectroscopy
avoids these challenges by using an on-chip microwave source and detector.
Usually there is no current through a Josephson junction voltage biased
below twice the superconducting gap. However, current peaks can develop
in its sub-gap \emph{$I\left(V\right)$} curve at voltages $2eV_{i}=h\nu_{i}$
if the energy $2eV$ lost by a Cooper pair tunneling across the barrier
can be absorbed as photons by the modes $\nu_{i}$ of the electromagnetic
environment of the junction, \textcolor{black}{like for example in
the case of Dynamical Coulomb Blockade} \citep{holst_effect_1994,hofheinz_bright_2011}.
By capacitively coupling an ancillary Josephson junction to the atomic-SQUID
(see Fig.~\ref{Fig: ICPT}) we expect to detect in the \emph{I-V}
of the junction the Andreev mode of the atomic contact. For the moment
we have simply tested this idea on a conventional two-junction SQUID
and detected, as shown in the right panel of Fig.~\ref{Fig: ICPT},
its plasma resonance corresponding to the oscillations of the LC circuit
formed by the Josephson inductance of the junctions and their intrinsic
capacitance.

\section{Conclusions}

Atomic-size contacts between metallic electrodes are now routinely
obtained by a variety of techniques. Because the number of conduction
channels they accomodate can be adjusted in-situ, and their corresponding
transmission coefficients are amenable to precise measurement, these
structures are model quantum point contacts which can be considered
as a test-bed for mesoscopic physics.

Besides the experiments described here, we have also tested quantitatively
the predictions for other transport phenomena: \textcolor{black}{shot-noise
in both the normal and superconducting states \citep{cron_multiple-charge-quanta_2001};
dynamical Coulomb blockade in the normal state \citep{cron_atomic_2001,cron_dynamical_2001};
the crossover between supercurrent and MAR dissipative current in
voltage-biased superconducting contacts \citep{chauvin_crossover_2007};
the response of superconducting contacts to microwave irradiation
\citep{chauvin_superconducting_2006}.} In all cases the agreement
between theory and experiment is remarkable and the ensemble of results
strongly supports the scattering theory of quantum transport.

In the superconducting state atomic contacts are the simplest possible
weak-links and perfectly illustrate the key role played by the Andreev
bound states in the Josephson effect. The Andreev states in a short
single channel Josephson structure constitute a two-level system \citep{ivanov_two-level_1999}
that has been proposed as the basis for a new class of superconducting
qubits \citep{zazunov_andreev_2003,zazunov_dynamics_2005,lantz_flux_2002,desposito_controlled_2001}.
What is particularly interesting and novel is that in contrast with
all other superconducting qubits based on Josephson junction circuits
\citep{wendin_quantum_2007} an Andreev qubit is a microscopic two
level system more like spin qubits in semiconducting quantum dots.
It can be viewed as a superconducting {}``quantum dot'' possibly
allowing manipulation of the spin degree of freedom of a single quasiparticle
\citep{chtchelkatchev_andreev_2003,michelsen_manipulation_2008}.
As reviewed in this work we have already characterized Andreev states
by measuring the current they carry but the coherence properties of
Andreev doublets are a key issue still to be addressed. The relaxation
time of the excited state and the dephasing time of a quantum superposition
of the two states have to be measured, understood and if possible
controlled.

\textbf{Acknowledgments}. We gratefully acknowledge the expert technical
support of Pief Orfila and Pascal Senat, the input and assistance
of other members of the Quantronics group (Michel Devoret, Patrice
Bertet and Denis Vion) and the crucial contributions of the graduate
students and postdocs who participated in this research effort in
the past: Elke Scheer, Ronald Cron, Peter vom Stein, Martin Chauvin,
Maria Luisa Della Rocca, Benjamin Huard, Maciej Zgirski and Quentin
Le Masne. We have enjoyed along the years fruitful collaborations
with Nicolás Agraït, Juan Carlos Cuevas, Alfredo Levy Yeyati, Alvaro
Martín Rodero, Gabino Rubio and Jan van Ruitenbeek, and greatly benefited
from discussions with Joachim Ankerhold, Sebastián Bergeret, Hermann
Grabert, John Martinis and Vitaly Shumeiko.


\begin{thebibliography}{58}
\providecommand{\natexlab}[1]{#1}
\providecommand{\url}[1]{\texttt{#1}}
\providecommand{\urlprefix}{URL }
\expandafter\ifx\csname urlstyle\endcsname\relax
  \providecommand{\doi}[1]{doi:\discretionary{}{}{}#1}\else
  \providecommand{\doi}[1]{doi:\discretionary{}{}{}\begingroup
  \urlstyle{rm}\url{#1}\endgroup}\fi
\providecommand{\bibinfo}[2]{#2}

\bibitem[{Webb et~al.(1985)Webb, Washburn, Umbach, and
  Laibowitz}]{webb_observation_1985}
\bibinfo{author}{R.~A. Webb}, \bibinfo{author}{S.~Washburn},
  \bibinfo{author}{C.~P. Umbach}, \bibinfo{author}{R.~B. Laibowitz},
  \bibinfo{title}{Observation of he {Aharonov-Bohm} Oscillations in
  normal-metal rings}, \bibinfo{journal}{Physical Review Letters}
  \bibinfo{volume}{54}~(\bibinfo{number}{25}) (\bibinfo{year}{1985})
  \bibinfo{pages}{2696}.

\bibitem[{Landauer(1970)}]{landauer_electrical_1970}
\bibinfo{author}{R.~Landauer}, \bibinfo{title}{Electrical resistance of
  disordered one-dimensional lattices}, \bibinfo{journal}{Philosophical
  Magazine} \bibinfo{volume}{21}~(\bibinfo{number}{172}) (\bibinfo{year}{1970})
  \bibinfo{pages}{863{\textendash}867}.
  
\bibitem[{Martin and Landauer(1992)}]{martin_wave-packet_1992}
\bibinfo{author}{T.~Martin}, \bibinfo{author}{R.~Landauer},
  \bibinfo{title}{Wave-packet approach to noise in multichannel mesoscopic
  systems}, \bibinfo{journal}{Physical Review B}
  \bibinfo{volume}{45}~(\bibinfo{number}{4}) (\bibinfo{year}{1992})
  \bibinfo{pages}{1742}.

\bibitem[{Nazarov and Blanter(2009)}]{nazarov_quantum_2009}
\bibinfo{author}{Y.~Nazarov}, \bibinfo{author}{Y.~M. Blanter},
  \bibinfo{title}{Quantum transport : introduction to nanoscience},
  \bibinfo{publisher}{Cambridge University Press}, \bibinfo{address}{Cambridge
  {UK;} New York}, \bibinfo{year}{2009}.

\bibitem[{van Wees et~al.(1988)van Wees, van Houten, Beenakker, Williamson,
  Kouwenhoven, van~der Marel, and Foxon}]{van_wees_quantized_1988}
\bibinfo{author}{B.~J. van Wees}, \bibinfo{author}{H.~van Houten},
  \bibinfo{author}{C.~W.~J. Beenakker}, \bibinfo{author}{J.~G. Williamson},
  \bibinfo{author}{L.~P. Kouwenhoven}, \bibinfo{author}{D.~van~der Marel},
  \bibinfo{author}{C.~T. Foxon}, \bibinfo{title}{Quantized conductance of point
  contacts in a two-dimensional electron gas}, \bibinfo{journal}{Physical
  Review Letters} \bibinfo{volume}{60}~(\bibinfo{number}{9})
  (\bibinfo{year}{1988}) \bibinfo{pages}{848}.

\bibitem[{Wharam et~al.(1988)Wharam, Thornton, Newbury, Pepper, Ahmed, Frost,
  Hasko, Peacock, Ritchie, and Jones}]{wharam_one-dimensional_1988}
\bibinfo{author}{D.~A. Wharam}, \bibinfo{author}{T.~J. Thornton},
  \bibinfo{author}{R.~Newbury}, \bibinfo{author}{M.~Pepper},
  \bibinfo{author}{H.~Ahmed}, \bibinfo{author}{J.~E.~F. Frost},
  \bibinfo{author}{D.~G. Hasko}, \bibinfo{author}{D.~C. Peacock},
  \bibinfo{author}{D.~A. Ritchie}, \bibinfo{author}{G.~A.~C. Jones},
  \bibinfo{title}{One-dimensional transport and the quantisation of the
  ballistic resistance}, \bibinfo{journal}{Journal of Physics C: Solid State
  Physics} \bibinfo{volume}{21}~(\bibinfo{number}{8}) (\bibinfo{year}{1988})
  \bibinfo{pages}{L209--L214}.

\bibitem[{Agra\"{i}t et~al.(2003)Agra\"{i}t, Levy~Yeyati, and van
  Ruitenbeek}]{agrait_quantum_2003}
\bibinfo{author}{N.~Agra\"{i}t}, \bibinfo{author}{A.~Levy~Yeyati},
  \bibinfo{author}{J.~M. van Ruitenbeek}, \bibinfo{title}{Quantum properties of
  atomic-sized conductors}, \bibinfo{journal}{Physics Reports}
  \bibinfo{volume}{377} (\bibinfo{year}{2003}) \bibinfo{pages}{81}.

\bibitem[{Eigler et~al.(1991)Eigler, Lutz, and Rudge}]{eigler_atomic_1991}
\bibinfo{author}{D.~M. Eigler}, \bibinfo{author}{C.~P. Lutz},
  \bibinfo{author}{W.~E. Rudge}, \bibinfo{title}{An atomic switch realized with
  the scanning tunnelling microscope}, \bibinfo{journal}{Nature}
  \bibinfo{volume}{352}~(\bibinfo{number}{6336}) (\bibinfo{year}{1991})
  \bibinfo{pages}{600{\textendash}603}.

\bibitem[{Moreland and Ekin(1985)}]{moreland_electron_1985}
\bibinfo{author}{J.~Moreland}, \bibinfo{author}{J.~W. Ekin},
  \bibinfo{title}{Electron tunneling experiments using {Nb-Sn}
  {\textquoteleft}{\textquoteleft}break{\textquoteright}{\textquoteright}
  junctions}, \bibinfo{journal}{Journal of Applied Physics}
  \bibinfo{volume}{58}~(\bibinfo{number}{10}) (\bibinfo{year}{1985})
  \bibinfo{pages}{3888}.

\bibitem[{Muller et~al.(1992)Muller, van Ruitenbeek, and
  de~Jongh}]{muller_experimental_1992}
\bibinfo{author}{C.~J. Muller}, \bibinfo{author}{J.~M. van Ruitenbeek},
  \bibinfo{author}{L.~J. de~Jongh}, \bibinfo{title}{Experimental observation of
  the transition from weak link to tunnel junction}, \bibinfo{journal}{Physica
  C: Superconductivity} \bibinfo{volume}{191}~(\bibinfo{number}{3-4})
  (\bibinfo{year}{1992}) \bibinfo{pages}{485{\textendash}504}.

\bibitem[{Scheer et~al.(1997)Scheer, Joyez, Esteve, Urbina, and
  Devoret}]{scheer_conduction_1997}
\bibinfo{author}{E.~Scheer}, \bibinfo{author}{P.~Joyez},
  \bibinfo{author}{D.~Esteve}, \bibinfo{author}{C.~Urbina},
  \bibinfo{author}{M.~H. Devoret}, \bibinfo{title}{Conduction channel
  transmissions of atomic-size aluminum contacts}, \bibinfo{journal}{Physical
  Review Letters} \bibinfo{volume}{78} (\bibinfo{year}{1997})
  \bibinfo{pages}{3535}.

\bibitem[{van Ruitenbeek et~al.(1996)van Ruitenbeek, Alvarez, Pi\~{n}eyro,
  Grahmann, Joyez, Devoret, Esteve, and
  Urbina}]{van_ruitenbeek_adjustable_1996}
\bibinfo{author}{J.~M. van Ruitenbeek}, \bibinfo{author}{A.~Alvarez},
  \bibinfo{author}{I.~Pi\~{n}eyro}, \bibinfo{author}{C.~Grahmann},
  \bibinfo{author}{P.~Joyez}, \bibinfo{author}{M.~H. Devoret},
  \bibinfo{author}{D.~Esteve}, \bibinfo{author}{C.~Urbina},
  \bibinfo{title}{Adjustable nanofabricated atomic size contacts},
  \bibinfo{journal}{Review of Scientific Instruments}
  \bibinfo{volume}{67}~(\bibinfo{number}{1}) (\bibinfo{year}{1996})
  \bibinfo{pages}{108}.

\bibitem[{Rubio et~al.(1996)Rubio, Agra\"{i}t, and
  Vieira}]{rubio_atomic-sized_1996}
\bibinfo{author}{G.~Rubio}, \bibinfo{author}{N.~Agra\"{i}t},
  \bibinfo{author}{S.~Vieira}, \bibinfo{title}{Atomic-sized Metallic Contacts:
  Mechanical Properties and Electronic Transport}, \bibinfo{journal}{Physical
  Review Letters} \bibinfo{volume}{76}~(\bibinfo{number}{13})
  (\bibinfo{year}{1996}) \bibinfo{pages}{2302}.

\bibitem[{{Rubio-Bollinger} et~al.(2004){Rubio-Bollinger}, Joyez, and
  Agra\"{i}t}]{rubio-bollinger_metallic_2004}
\bibinfo{author}{G.~{Rubio-Bollinger}}, \bibinfo{author}{P.~Joyez},
  \bibinfo{author}{N.~Agra\"{i}t}, \bibinfo{title}{Metallic Adhesion in
  {Atomic-Size} Junctions}, \bibinfo{journal}{Physical Review Letters}
  \bibinfo{volume}{93}~(\bibinfo{number}{11}) (\bibinfo{year}{2004})
  \bibinfo{pages}{116803}.

\bibitem[{Todorov and Sutton(1993)}]{todorov_jumps_1993}
\bibinfo{author}{T.~N. Todorov}, \bibinfo{author}{A.~P. Sutton},
  \bibinfo{title}{Jumps in electronic conductance due to mechanical
  instabilities}, \bibinfo{journal}{Physical Review Letters}
  \bibinfo{volume}{70}~(\bibinfo{number}{14}) (\bibinfo{year}{1993})
  \bibinfo{pages}{2138}.

\bibitem[{Brandbyge et~al.(1995)Brandbyge, Schiotz, Sorensen, Stoltze,
  Jacobsen, Norskov, Olesen, Laegsgaard, Stensgaard, and
  Besenbacher}]{brandbyge_quantized_1995}
\bibinfo{author}{M.~Brandbyge}, \bibinfo{author}{J.~Schiotz},
  \bibinfo{author}{M.~R. Sorensen}, \bibinfo{author}{P.~Stoltze},
  \bibinfo{author}{K.~W. Jacobsen}, \bibinfo{author}{J.~K. Norskov},
  \bibinfo{author}{L.~Olesen}, \bibinfo{author}{E.~Laegsgaard},
  \bibinfo{author}{I.~Stensgaard}, \bibinfo{author}{F.~Besenbacher},
  \bibinfo{title}{Quantized conductance in atom-sized wires between two
  metals}, \bibinfo{journal}{Physical Review B}
  \bibinfo{volume}{52}~(\bibinfo{number}{11}) (\bibinfo{year}{1995})
  \bibinfo{pages}{8499}.

\bibitem[{Sorensen et~al.(1998)Sorensen, Brandbyge, and
  Jacobsen}]{sorensen_mechanical_1998}
\bibinfo{author}{M.~R. Sorensen}, \bibinfo{author}{M.~Brandbyge},
  \bibinfo{author}{K.~W. Jacobsen}, \bibinfo{title}{Mechanical deformation of
  atomic-scale metallic contacts: Structure and mechanisms},
  \bibinfo{journal}{Physical Review B}
  \bibinfo{volume}{57}~(\bibinfo{number}{6}) (\bibinfo{year}{1998})
  \bibinfo{pages}{3283}.

\bibitem[{Yanson et~al.(1998)Yanson, Bollinger, van~den Brom, Agra\"{i}t, and
  van Ruitenbeek}]{yanson_formation_1998}
\bibinfo{author}{A.~I. Yanson}, \bibinfo{author}{G.~R. Bollinger},
  \bibinfo{author}{H.~E. van~den Brom}, \bibinfo{author}{N.~Agra\"{i}t},
  \bibinfo{author}{J.~M. van Ruitenbeek}, \bibinfo{title}{Formation and
  manipulation of a metallic wire of single gold atoms},
  \bibinfo{journal}{Nature} \bibinfo{volume}{395}~(\bibinfo{number}{6704})
  (\bibinfo{year}{1998}) \bibinfo{pages}{783{\textendash}785}.

\bibitem[{Ohnishi et~al.(1998)Ohnishi, Kondo, and
  Takayanagi}]{ohnishi_quantized_1998}
\bibinfo{author}{H.~Ohnishi}, \bibinfo{author}{Y.~Kondo},
  \bibinfo{author}{K.~Takayanagi}, \bibinfo{title}{Quantized conductance
  through individual rows of suspended gold atoms}, \bibinfo{journal}{Nature}
  \bibinfo{volume}{395}~(\bibinfo{number}{6704}) (\bibinfo{year}{1998})
  \bibinfo{pages}{780{\textendash}783}.

\bibitem[{Rodrigues and Ugarte(2001)}]{rodrigues_real-time_2001}
\bibinfo{author}{V.~Rodrigues}, \bibinfo{author}{D.~Ugarte},
  \bibinfo{title}{Real-time imaging of atomistic process in one-atom-thick
  metal junctions}, \bibinfo{journal}{Physical Review B}
  \bibinfo{volume}{63}~(\bibinfo{number}{7}) (\bibinfo{year}{2001})
  \bibinfo{pages}{073405}.

\bibitem[{Masuda et~al.(2009)Masuda, Monna, Matsuda, and Kizuka}]{masuda__2009}
\bibinfo{author}{H.~Masuda}, \bibinfo{author}{K.~Monna},
  \bibinfo{author}{T.~Matsuda}, \bibinfo{author}{T.~Kizuka},
  \bibinfo{journal}{{e-Journal} of Surface Science and Nanotechnology}
  \bibinfo{volume}{7} (\bibinfo{year}{2009})
  \bibinfo{pages}{549{\textendash}552}.

\bibitem[{Cron(2001)}]{cron_atomic_2001}
\bibinfo{author}{R.~Cron}, \bibinfo{title}{Atomic Contacts: a {Test-Bed} for
  Mesoscopic Physics}, \bibinfo{type}{{Ph.D.} thesis},
  \bibinfo{school}{Universit\'{e} Pierre et Marie Curie, Paris, France},
  \bibinfo{note}{available at
  http://tel.archives-ouvertes.fr/tel-00001329/fr/}, \bibinfo{year}{2001}.

\bibitem[{Klapwijk et~al.(1982)Klapwijk, Blonder, and
  Tinkham}]{klapwijk_explanation_1982}
\bibinfo{author}{T.~M. Klapwijk}, \bibinfo{author}{G.~E. Blonder},
  \bibinfo{author}{M.~Tinkham}, \bibinfo{title}{Explanation of subharmonic
  energy gap structure in superconducting contacts}, \bibinfo{journal}{Physica
  {B+C}} \bibinfo{volume}{109-110} (\bibinfo{year}{1982})
  \bibinfo{pages}{1657{\textendash}1664}.

\bibitem[{Blonder et~al.(1982)Blonder, Tinkham, and
  Klapwijk}]{blonder_transition_1982}
\bibinfo{author}{G.~E. Blonder}, \bibinfo{author}{M.~Tinkham},
  \bibinfo{author}{T.~M. Klapwijk}, \bibinfo{title}{Transition from metallic to
  tunneling regimes in superconducting microconstrictions: Excess current,
  charge imbalance, and supercurrent conversion}, \bibinfo{journal}{Physical
  Review B} \bibinfo{volume}{25}~(\bibinfo{number}{7}) (\bibinfo{year}{1982})
  \bibinfo{pages}{4515}.

\bibitem[{Arnold(1987)}]{arnold_superconducting_1987}
\bibinfo{author}{G.~B. Arnold}, \bibinfo{title}{Superconducting tunneling
  without the tunneling Hamiltonian. {II.} Subgap harmonic structure},
  \bibinfo{journal}{Journal of Low Temperature Physics} \bibinfo{volume}{68}
  (\bibinfo{year}{1987}) \bibinfo{pages}{1}.

\bibitem[{Averin and Bardas(1995)}]{averin_ac_1995}
\bibinfo{author}{D.~Averin}, \bibinfo{author}{A.~Bardas}, \bibinfo{title}{ac
  Josephson Effect in a Single Quantum Channel}, \bibinfo{journal}{Physical
  Review Letters} \bibinfo{volume}{75}~(\bibinfo{number}{9})
  (\bibinfo{year}{1995}) \bibinfo{pages}{1831}.

\bibitem[{Cuevas et~al.(1996)Cuevas, {Mart\'{i}n-Rodero}, and
  Yeyati}]{cuevas_hamiltonian_1996}
\bibinfo{author}{J.~C. Cuevas}, \bibinfo{author}{A.~{Mart\'{i}n-Rodero}},
  \bibinfo{author}{A.~L. Yeyati}, \bibinfo{title}{Hamiltonian approach to the
  transport properties of superconducting quantum point contacts},
  \bibinfo{journal}{Physical Review B}
  \bibinfo{volume}{54}~(\bibinfo{number}{10}) (\bibinfo{year}{1996})
  \bibinfo{pages}{7366}.

\bibitem[{Riquelme et~al.(2005)Riquelme, Vega, Yeyati, Agra\"{i}t,
  {Martin-Rodero}, and {Rubio-Bollinger}}]{riquelme_distribution_2005}
\bibinfo{author}{J.~J. Riquelme}, \bibinfo{author}{L.~d.~l. Vega},
  \bibinfo{author}{A.~L. Yeyati}, \bibinfo{author}{N.~Agra\"{i}t},
  \bibinfo{author}{A.~{Martin-Rodero}}, \bibinfo{author}{G.~{Rubio-Bollinger}},
  \bibinfo{title}{Distribution of conduction channels in nanoscale contacts:
  Evolution towards the diffusive limit}, \bibinfo{journal}{Europhysics Letters
  {(EPL)}} \bibinfo{volume}{70}~(\bibinfo{number}{5}) (\bibinfo{year}{2005})
  \bibinfo{pages}{663{\textendash}669}.

\bibitem[{Scheer et~al.(1998)Scheer, Agra\"{i}t, Cuevas, Levy~Yeyati, Ludoph,
  {Mart\'{i}n-Rodero}, Rubio~Bollinger, van Ruitenbeek, and
  Urbina}]{scheer_signature_1998}
\bibinfo{author}{E.~Scheer}, \bibinfo{author}{N.~Agra\"{i}t},
  \bibinfo{author}{J.~C. Cuevas}, \bibinfo{author}{A.~Levy~Yeyati},
  \bibinfo{author}{B.~Ludoph}, \bibinfo{author}{A.~{Mart\'{i}n-Rodero}},
  \bibinfo{author}{G.~Rubio~Bollinger}, \bibinfo{author}{J.~M. van Ruitenbeek},
  \bibinfo{author}{C.~Urbina}, \bibinfo{title}{The signature of chemical
  valence in the electrical conduction through a single-atom contact},
  \bibinfo{journal}{Nature {(London)}} \bibinfo{volume}{394}
  (\bibinfo{year}{1998}) \bibinfo{pages}{154}.

\bibitem[{Scheer et~al.(2001)Scheer, Belzig, Naveh, Devoret, Esteve, and
  Urbina}]{scheer_proximity_2001}
\bibinfo{author}{E.~Scheer}, \bibinfo{author}{W.~Belzig},
  \bibinfo{author}{Y.~Naveh}, \bibinfo{author}{M.~H. Devoret},
  \bibinfo{author}{D.~Esteve}, \bibinfo{author}{C.~Urbina},
  \bibinfo{title}{Proximity Effect and Multiple Andreev Reflections in Gold
  Atomic Contacts}, \bibinfo{journal}{Physical Review Letters}
  \bibinfo{volume}{86}~(\bibinfo{number}{2}) (\bibinfo{year}{2001})
  \bibinfo{pages}{284}.

\bibitem[{Josephson(1962)}]{josephson_possible_1962}
\bibinfo{author}{B.~D. Josephson}, \bibinfo{title}{Possible new effects in
  superconductive tunnelling}, \bibinfo{journal}{Phys. Lett.}
  \bibinfo{volume}{1} (\bibinfo{year}{1962}) \bibinfo{pages}{251}.

\bibitem[{Golubov et~al.(2004)Golubov, Kupriyanov, and
  Il'ichev}]{golubov_current-phase_2004}
\bibinfo{author}{A.~A. Golubov}, \bibinfo{author}{M.~Y. Kupriyanov},
  \bibinfo{author}{E.~Il'ichev}, \bibinfo{title}{The current-phase relation in
  Josephson junctions}, \bibinfo{journal}{Reviews of Modern Physics}
  \bibinfo{volume}{76}~(\bibinfo{number}{2}) (\bibinfo{year}{2004})
  \bibinfo{pages}{411}.

\bibitem[{Furusaki and Tsukada(1990)}]{furusaki_unified_1990}
\bibinfo{author}{A.~Furusaki}, \bibinfo{author}{M.~Tsukada}, \bibinfo{title}{A
  unified theory of clean Josephson junctions}, \bibinfo{journal}{Physica B:
  Condensed Matter} \bibinfo{volume}{165-166}~(\bibinfo{number}{Part 2})
  (\bibinfo{year}{1990}) \bibinfo{pages}{967{\textendash}968}.

\bibitem[{Beenakker and van Houten(1991)}]{beenakker_josephson_1991}
\bibinfo{author}{C.~W.~J. Beenakker}, \bibinfo{author}{H.~van Houten},
  \bibinfo{title}{Josephson current through a superconducting quantum point
  contact shorter than the coherence length}, \bibinfo{journal}{Physical Review
  Letters} \bibinfo{volume}{66}~(\bibinfo{number}{23}) (\bibinfo{year}{1991})
  \bibinfo{pages}{3056}.

\bibitem[{Andreev(1964)}]{andreev__1964}
\bibinfo{author}{A.~F. Andreev}, \bibinfo{journal}{Sov. Phys. {JETP}}
  \bibinfo{volume}{19} (\bibinfo{year}{1964}) \bibinfo{pages}{1228}.

\bibitem[{Bagwell(1992)}]{bagwell_suppression_1992}
\bibinfo{author}{P.~F. Bagwell}, \bibinfo{title}{Suppression of the Josephson
  current through a narrow, mesoscopic, semiconductor channel by a single
  impurity}, \bibinfo{journal}{Phys. Rev. B}
  \bibinfo{volume}{46}~(\bibinfo{number}{19}) (\bibinfo{year}{1992})
  \bibinfo{pages}{12573{\textendash}12586}.

\bibitem[{Chtchelkatchev and Nazarov(2003)}]{chtchelkatchev_andreev_2003}
\bibinfo{author}{N.~M. Chtchelkatchev}, \bibinfo{author}{Y.~V. Nazarov},
  \bibinfo{title}{Andreev Quantum Dots for Spin Manipulation},
  \bibinfo{journal}{Physical Review Letters}
  \bibinfo{volume}{90}~(\bibinfo{number}{22}) (\bibinfo{year}{2003})
  \bibinfo{pages}{226806}.

\bibitem[{Goffman et~al.(2000)Goffman, Cron, Levy~Yeyati, Joyez, Devoret,
  Esteve, and Urbina}]{goffman_supercurrent_2000}
\bibinfo{author}{M.~F. Goffman}, \bibinfo{author}{R.~Cron},
  \bibinfo{author}{A.~Levy~Yeyati}, \bibinfo{author}{P.~Joyez},
  \bibinfo{author}{M.~H. Devoret}, \bibinfo{author}{D.~Esteve},
  \bibinfo{author}{C.~Urbina}, \bibinfo{title}{Supercurrent in Atomic Point
  Contacts and Andreev States}, \bibinfo{journal}{Physical Review Letters}
  \bibinfo{volume}{85}~(\bibinfo{number}{1}) (\bibinfo{year}{2000})
  \bibinfo{pages}{170}.

\bibitem[{Della~Rocca et~al.(2007)Della~Rocca, Chauvin, Huard, Pothier, Esteve,
  and Urbina}]{della_rocca_measurement_2007}
\bibinfo{author}{M.~L. Della~Rocca}, \bibinfo{author}{M.~Chauvin},
  \bibinfo{author}{B.~Huard}, \bibinfo{author}{H.~Pothier},
  \bibinfo{author}{D.~Esteve}, \bibinfo{author}{C.~Urbina},
  \bibinfo{title}{Measurement of the {Current-Phase} Relation of
  Superconducting Atomic Contacts}, \bibinfo{journal}{Physical Review Letters}
  \bibinfo{volume}{99}~(\bibinfo{number}{12}) (\bibinfo{year}{2007})
  \bibinfo{pages}{127005{\textendash}4}.

\bibitem[{Huard(2006)}]{huard_interactions_2006}
\bibinfo{author}{B.~Huard}, \bibinfo{title}{Interactions between electrons,
  mesoscopic Josephson effect and asymmetric current fluctuations}, Ph.D.
  thesis, \bibinfo{school}{Universit\'{e} Pierre et Marie Curie, Paris,
  France},
  \bibinfo{note}{available at
  http://tel.archives-ouvertes.fr/tel-00119371/fr/}, \bibinfo{year}{2006}.

\bibitem[{Chauvin(2005)}]{chauvin_effet_2005}
\bibinfo{author}{M.~Chauvin}, \bibinfo{title}{Effet Josephson dans les contacts
  atomiques / The Josephson Effect in Atomic Contacts}, \bibinfo{type}{{Ph.D.}
  thesis}, \bibinfo{school}{Universit\'{e} Pierre et Marie Curie, Paris,
  France}, \bibinfo{note}{available
  at http://tel.archives-ouvertes.fr/tel-00107465/fr/}, \bibinfo{year}{2005}.

\bibitem[{{Lefevre-Seguin} et~al.(1992){Lefevre-Seguin}, Turlot, Urbina,
  Esteve, and Devoret}]{lefevre-seguin_thermal_1992}
\bibinfo{author}{V.~{Lefevre-Seguin}}, \bibinfo{author}{E.~Turlot},
  \bibinfo{author}{C.~Urbina}, \bibinfo{author}{D.~Esteve},
  \bibinfo{author}{M.~H. Devoret}, \bibinfo{title}{Thermal activation of a
  hysteretic dc superconducting quantum interference device from its different
  zero-voltage states}, \bibinfo{journal}{Physical Review B {(Condensed} Matter
  and Materials Physics)} \bibinfo{volume}{46}~(\bibinfo{number}{9})
  (\bibinfo{year}{1992}) \bibinfo{pages}{5507{\textendash}5522}.

\bibitem[{Le~Masne(2009)}]{le_masne_asymmetric_2009}
\bibinfo{author}{Q.~Le~Masne}, \bibinfo{title}{Asymmetric current fluctuations
  and Andreev states probed with a Josephson junction}, Ph.D. thesis,
  \bibinfo{school}{Universit\'{e} Pierre et Marie Curie, Paris, France},
  \bibinfo{note}{available at
  http://tel.archives-ouvertes.fr/tel-00482483/fr/}, \bibinfo{year}{2009}.

\bibitem[{Desp\'{o}sito and Levy~Yeyati(2001)}]{desposito_controlled_2001}
\bibinfo{author}{M.~A. Desp\'{o}sito}, \bibinfo{author}{A.~Levy~Yeyati},
  \bibinfo{title}{Controlled dephasing of Andreev states in superconducting
  quantum point contacts}, \bibinfo{journal}{Physical Review B}
  \bibinfo{volume}{64}~(\bibinfo{number}{14}) (\bibinfo{year}{2001})
  \bibinfo{pages}{140511}.

\bibitem[{Zgirski et~al.(2011)Zgirski, Bretheau, Le~Masne, Pothier, Esteve, and
  Urbina}]{zgirski_evidence_2011}
\bibinfo{author}{M.~Zgirski}, \bibinfo{author}{L.~Bretheau},
  \bibinfo{author}{Q.~Le~Masne}, \bibinfo{author}{H.~Pothier},
  \bibinfo{author}{D.~Esteve}, \bibinfo{author}{C.~Urbina},
  \bibinfo{title}{Evidence for long-lived quasiparticles Trapped in
  Superconducting Point Contacts}, \bibinfo{journal}{Physical Review Letters}
  \bibinfo{volume}{106}~(\bibinfo{number}{25}) (\bibinfo{year}{2011})
  \bibinfo{pages}{257003}.

\bibitem[{Blais et~al.(2004)Blais, Huang, Wallraff, Girvin, and
  Schoelkopf}]{blais_cavity_2004}
\bibinfo{author}{A.~Blais}, \bibinfo{author}{R.~Huang},
  \bibinfo{author}{A.~Wallraff}, \bibinfo{author}{S.~M. Girvin},
  \bibinfo{author}{R.~J. Schoelkopf}, \bibinfo{title}{Cavity quantum
  electrodynamics for superconducting electrical circuits: An architecture for
  quantum computation}, \bibinfo{journal}{Physical Review A}
  \bibinfo{volume}{69} (\bibinfo{year}{2004}) \bibinfo{pages}{62320}.

\bibitem[{Holst et~al.(1994)Holst, Esteve, Urbina, and
  Devoret}]{holst_effect_1994}
\bibinfo{author}{T.~Holst}, \bibinfo{author}{D.~Esteve},
  \bibinfo{author}{C.~Urbina}, \bibinfo{author}{M.~H. Devoret},
  \bibinfo{title}{Effect of a Transmission Line Resonator on a Small
  Capacitance Tunnel Junction}, \bibinfo{journal}{Physical Review Letters}
  \bibinfo{volume}{73}~(\bibinfo{number}{25}) (\bibinfo{year}{1994})
  \bibinfo{pages}{3455{\textendash}3458}.

\bibitem[{Hofheinz et~al.(2011)Hofheinz, Portier, Baudouin, Joyez, Vion,
  Bertet, Roche, and Esteve}]{hofheinz_bright_2011}
\bibinfo{author}{M.~Hofheinz}, \bibinfo{author}{F.~Portier},
  \bibinfo{author}{Q.~Baudouin}, \bibinfo{author}{P.~Joyez},
  \bibinfo{author}{D.~Vion}, \bibinfo{author}{P.~Bertet},
  \bibinfo{author}{P.~Roche}, \bibinfo{author}{D.~Esteve},
  \bibinfo{title}{Bright Side of the Coulomb Blockade},
  \bibinfo{journal}{Physical Review Letters}
  \bibinfo{volume}{106}~(\bibinfo{number}{21}) (\bibinfo{year}{2011})
  \bibinfo{pages}{217005}.

\bibitem[{Cron et~al.(2001{\natexlab{a}})Cron, Goffman, Esteve, and
  Urbina}]{cron_multiple-charge-quanta_2001}
\bibinfo{author}{R.~Cron}, \bibinfo{author}{M.~F. Goffman},
  \bibinfo{author}{D.~Esteve}, \bibinfo{author}{C.~Urbina},
  \bibinfo{title}{{Multiple-Charge-Quanta} Shot Noise in Superconducting Atomic
  Contacts}, \bibinfo{journal}{Physical Review Letters}
  \bibinfo{volume}{86}~(\bibinfo{number}{18})
  (\bibinfo{year}{2001}{\natexlab{a}}) \bibinfo{pages}{4104{\textendash}4107}.

\bibitem[{Cron et~al.(2001{\natexlab{b}})Cron, Vecino, Devoret, Esteve, Joyez,
  Levy~Yeyati, {Martin-Rodero}, and Urbina}]{cron_dynamical_2001}
\bibinfo{author}{R.~Cron}, \bibinfo{author}{E.~Vecino}, \bibinfo{author}{M.~H.
  Devoret}, \bibinfo{author}{D.~Esteve}, \bibinfo{author}{P.~Joyez},
  \bibinfo{author}{A.~Levy~Yeyati}, \bibinfo{author}{A.~{Martin-Rodero}},
  \bibinfo{author}{C.~Urbina}, \bibinfo{title}{Dynamical Coulomb blockade in
  quantum point contacts}, in: \bibinfo{booktitle}{Electronic Correlations:
  from meso- to nano- physics}, \bibinfo{publisher}{{EDP} Sciences},
  \bibinfo{year}{2001}{\natexlab{b}}.

\bibitem[{Chauvin et~al.(2007)Chauvin, vom Stein, Esteve, Urbina, Cuevas, and
  Yeyati}]{chauvin_crossover_2007}
\bibinfo{author}{M.~Chauvin}, \bibinfo{author}{P.~vom Stein},
  \bibinfo{author}{D.~Esteve}, \bibinfo{author}{C.~Urbina},
  \bibinfo{author}{J.~C. Cuevas}, \bibinfo{author}{A.~L. Yeyati},
  \bibinfo{title}{Crossover from Josephson to Multiple Andreev Reflection
  Currents in Atomic Contacts}, \bibinfo{journal}{Physical Review Letters}
  \bibinfo{volume}{99}~(\bibinfo{number}{6}) (\bibinfo{year}{2007})
  \bibinfo{pages}{067008}.

\bibitem[{Chauvin et~al.(2006)Chauvin, Stein, Pothier, Joyez, Huber, Esteve,
  and Urbina}]{chauvin_superconducting_2006}
\bibinfo{author}{M.~Chauvin}, \bibinfo{author}{P.~v. Stein},
  \bibinfo{author}{H.~Pothier}, \bibinfo{author}{P.~Joyez},
  \bibinfo{author}{M.~E. Huber}, \bibinfo{author}{D.~Esteve},
  \bibinfo{author}{C.~Urbina}, \bibinfo{title}{Superconducting Atomic Contacts
  under Microwave Irradiation}, \bibinfo{journal}{Physical Review Letters}
  \bibinfo{volume}{97}~(\bibinfo{number}{6}) (\bibinfo{year}{2006})
  \bibinfo{pages}{067006}.

\bibitem[{Ivanov and Feigel'man(1999)}]{ivanov_two-level_1999}
\bibinfo{author}{D.~A. Ivanov}, \bibinfo{author}{M.~V. Feigel'man},
  \bibinfo{title}{Two-level Hamiltonian of a superconducting quantum point
  contact}, \bibinfo{journal}{Phys. Rev. B}
  \bibinfo{volume}{59}~(\bibinfo{number}{13}) (\bibinfo{year}{1999})
  \bibinfo{pages}{8444{\textendash}8446}.

\bibitem[{Zazunov et~al.(2003)Zazunov, Shumeiko, Bratus', Lantz, and
  Wendin}]{zazunov_andreev_2003}
\bibinfo{author}{A.~Zazunov}, \bibinfo{author}{V.~S. Shumeiko},
  \bibinfo{author}{E.~N. Bratus'}, \bibinfo{author}{J.~Lantz},
  \bibinfo{author}{G.~Wendin}, \bibinfo{title}{Andreev Level Qubit},
  \bibinfo{journal}{Physical Review Letters}
  \bibinfo{volume}{90}~(\bibinfo{number}{8}) (\bibinfo{year}{2003})
  \bibinfo{pages}{087003}.

\bibitem[{Zazunov et~al.(2005)Zazunov, Shumeiko, Wendin, and
  Bratus'}]{zazunov_dynamics_2005}
\bibinfo{author}{A.~Zazunov}, \bibinfo{author}{V.~S. Shumeiko},
  \bibinfo{author}{G.~Wendin}, \bibinfo{author}{E.~N. Bratus'},
  \bibinfo{title}{Dynamics and phonon-induced decoherence of Andreev level
  qubit}, \bibinfo{journal}{Physical Review B {(Condensed} Matter and Materials
  Physics)} \bibinfo{volume}{71}~(\bibinfo{number}{21}) (\bibinfo{year}{2005})
  \bibinfo{pages}{214505}.

\bibitem[{Lantz et~al.(2002)Lantz, Shumeiko, Bratus, and
  Wendin}]{lantz_flux_2002}
\bibinfo{author}{J.~Lantz}, \bibinfo{author}{V.~S. Shumeiko},
  \bibinfo{author}{E.~Bratus}, \bibinfo{author}{G.~Wendin},
  \bibinfo{title}{Flux qubit with a quantum point contact},
  \bibinfo{journal}{Physica C: Superconductivity} \bibinfo{volume}{368}
  (\bibinfo{year}{2002}) \bibinfo{pages}{315}.

\bibitem[{Wendin and Shumeiko(2007)}]{wendin_quantum_2007}
\bibinfo{author}{G.~Wendin}, \bibinfo{author}{V.~S. Shumeiko},
  \bibinfo{title}{Quantum bits with Josephson junctions {(Review} Article)},
  \bibinfo{journal}{Low Temperature Physics}
  \bibinfo{volume}{33}~(\bibinfo{number}{9}) (\bibinfo{year}{2007})
  \bibinfo{pages}{724}.

\bibitem[{Michelsen et~al.(2008)Michelsen, Shumeiko, and
  Wendin}]{michelsen_manipulation_2008}
\bibinfo{author}{J.~Michelsen}, \bibinfo{author}{V.~S. Shumeiko},
  \bibinfo{author}{G.~Wendin}, \bibinfo{title}{Manipulation with Andreev states
  in spin active mesoscopic Josephson junctions}, \bibinfo{journal}{Physical
  Review B} \bibinfo{volume}{77}~(\bibinfo{number}{18}) (\bibinfo{year}{2008})
  \bibinfo{pages}{184506}.

\end{thebibliography}
\end{document}